\documentclass[aps,twocolumn,amsfonts,nofootinbib,preprintnumbers,eqsecnum,superscriptaddress]{revtex4-1}

\usepackage{epsfig} 
\usepackage{psfrag}
\usepackage{amsmath}
\usepackage{amssymb}
\usepackage{color}
\usepackage{graphicx}
\usepackage{subfigure}
\usepackage{hyperref} 
\usepackage{rotating}
\usepackage{enumerate,mdwlist}

\def\be{\begin{equation}}
\def\ee{\end{equation}}
\newcommand{\js}[1]{{\color{red}{[--Julian: {#1}]}}}

\def\bea{\begin{eqnarray}}
\def\eea{\end{eqnarray}}

\newcommand{\nb}[1]{\color{blue}}

\newcommand{\hl}[1]{\color{magenta}}

\newcommand\ov{\over}

\usepackage[normalem]{ulem}
\usepackage{amsfonts}
\usepackage{epsfig}
\usepackage{mathbbol}

\usepackage{amsfonts}

\def\Tr{\mathop{\rm Tr}}

\newcommand\p{\ensuremath{\partial}}

\newcommand\field[1]{{\ensuremath{\mathbb{{#1}}}}}

\newcommand\vev[1]{{\ensuremath{\left\langle{#1}\right\rangle}}}

\newcommand\ket[1]{\ensuremath{\lvert{#1}\rangle}}

\newcommand{\RR}{\field{R}}

\newcommand{\bi}{\begin{itemize}}
\newcommand{\ei}{\end{itemize}}
\newcommand{\ben}{\begin{enumerate}}
\newcommand{\een}{\end{enumerate}}
\newcommand{\bca}{\begin{cases}}
\newcommand{\eca}{\end{cases}}
\newcommand{\bln}{\begin{align}}
\newcommand{\eln}{\end{align}}
\newcommand{\bst}{\begin{split}}
\newcommand{\est}{\end{split}}
\def\ie{\begin{equation}\begin{aligned}}
\def\fe{\end{aligned}\end{equation}}

\newcommand\al{{\alpha}}

\newcommand\sig{\sigma}

\newcommand\lam{\lambda}

\newcommand\om{\omega}
\newcommand\Om{\Omega}

\newcommand\ga{{\ensuremath{{\gamma}}}}

\newcommand\De{{\ensuremath{{\Delta}}}}

\def\le{\left}
\def\ri{\right}

\newcommand\sD{{\ensuremath{{\mathcal D}}}}
\newcommand\sE{{\ensuremath{{\mathcal E}}}}

\newcommand\sI{{\ensuremath{{\mathcal I}}}}

\newcommand\sH{{\ensuremath{{\mathcal H}}}}

\newcommand\sN{{\ensuremath{{\mathcal N}}}}
\newcommand\sO{{\ensuremath{{\mathcal O}}}}
\newcommand\sV{{\mathcal V}}

\newcommand\sR{{\mathcal R}}
\newcommand\sS{{\mathcal S}}

\newcommand\vx{{\vec x}}

\newcommand{\fa}{{\mathfrak a}}

\begin{document}

\preprint{MIT-CTP/5194}

\title{Quantum many-body physics from a gravitational lens}
  \author{Hong Liu}
    \affiliation{Center for theoretical physics, Massachusetts Institute of Technology, Cambridge, MA 02139, U.S.A.}
 \author{Julian Sonner}
    \affiliation{Department of theoretical physics, Universit\'e de Gen\`eve, 24 quai Ernest-Ansermet, 1211 Gen\`eve 4, Switzerland}
\begin{abstract}

The last two decades have seen the emergence of stunning interconnections among various previously remotely related disciplines such as condensed matter, nuclear physics, gravity and quantum information, fueled both by experimental advances and new powerful theoretical methods brought by holographic duality. In this non-technical review we sample some recent developments in holographic duality in connection 
with quantum many-body dynamics. These include insights into strongly correlated phases 
without quasiparticles and their transport properties, quantum many-body chaos, and scrambling of quantum 
information. We also discuss recent progress in understanding the structure of holographic duality itself using quantum information, including a ``local'' version of the duality as well as the
quantum error correction interpretation of quantum many-body states with a gravity dual, and how such notions help demonstrate the unitarity of black hole evaporation.

\end{abstract}

\date{\today}
\maketitle
\tableofcontents

\section{Strongly correlated systems and black holes}

Traditionally, gravitational, nuclear, and condensed matter physics are regarded as remotely-connected disciplines with completely different sets of goals and physical laws. Remarkably, during the last fifteen years, stunning interconnections have emerged among them, with the unifying theme that {\it strongly correlated quantum liquids without quasiparticles} often exhibit universal behavior which does not depend on the details of its constituents.  One engine for this ``unification'' among different disciplines has been an accumulation of experimental results, in strongly correlated electronic systems such as high temperature superconducting cuprates, in the quark-gluon plasmas (QGP) created at Relativistic Heavy Ion Collider (RHIC) and Large Hadron Collider (LHC), and in ultracold quantum gases. Another engine are new powerful theoretical methods brought by holographic duality~\cite{Maldacena:1997re,Gubser:1998bc,Witten:1998qj} and quantum information.

If interactions among the constituent particles of a quantum many-body system are weak--with the potential energy being much smaller than the kinetic energy of a particle--the effects of interactions can be treated as small perturbations of an ideal gas. Surprisingly, even when the constituent particles interact strongly at a microscopic level, a weakly interacting picture often  emerges macroscopically. For example, in most metallic systems, while the underlying Coulomb interactions among electrons can be strong, macroscopic properties are controlled by long-lived collective particle-like excitations which carry the same charge and spin as electrons, but interact weakly with one another. Powerful techniques of perturbation theory and the Boltzmann equations can then be applied to the effective theory of these ``quasi-electrons''
to obtain thermodynamic and transport properties. This {\it quasiparticle}  paradigm, namely the existence of long-lived particle-like excitations which control macroscopic properties, has been a cornerstone of condensed matter physics and extremely successful in explaining a wide variety of phenomena, from metallic properties to magnetism to superfluidity and superconductivity. 

 A defining feature of systems with a quasiparticle description is a separation of scales, with the scattering rate among quasiparticles (which characterizes the strength of interactions) much smaller than the temperature $T$ (which characterizes typical kinetic energies), i.e.  ${\hbar \ov \tau} \ll {T}$ ($\tau$ is the scattering time). 

In various strongly correlated quantum liquids, such as 
the normal state of a high-temperature superconducting cuprate, the QGP at RHIC and LHC, and ultracold atomic systems at unitarity, such a separation of scales is absent. While these systems operate at vast different scales with completely different underlying interactions,  they all exhibit so-called  ``Planckian dissipation''~\cite{ZaanenPlanckDiss,sachdev_2011}--when trying to fit transport properties of these systems into the quasiparticle paradigm, one finds that the scattering time among ``would-be'' quasiparticles 
scales as $\tau \sim \frac{\hbar}{k_B T}$.  
For such systems the quasiparticle paradigm cannot be self-consistently applied as the lifetime of a ``would-be'' quasiparticle 
is too short to have any long-term effects.
Indeed traditional methods based on perturbation theory and Boltzmann equations fail dramatically to explain their thermodynamic and transport properties. Finding a general framework to treat quantum liquids without quasiparticles 
is one of the most outstanding theoretical challenges and has far-reaching implications for diverse disciplines.

For a microscopically strongly interacting system then the question arises as to what determines the emergence or lack of a quasiparticle description macroscopically. The answer to this question is also not understood. There are, however, indications that the quantum informational structure of the ground state and/or low-lying states may play a key role, as systems with a weakly coupled quasiparticle 
description appear to only exhibit short-range entanglement. %(\HL{More?}) \js{I dont' think much more can be said.}

%is goverened by so-called ``Planckian dissipation,'' 

\begin{figure}[t]
\begin{center}
\includegraphics[width=.9\columnwidth]{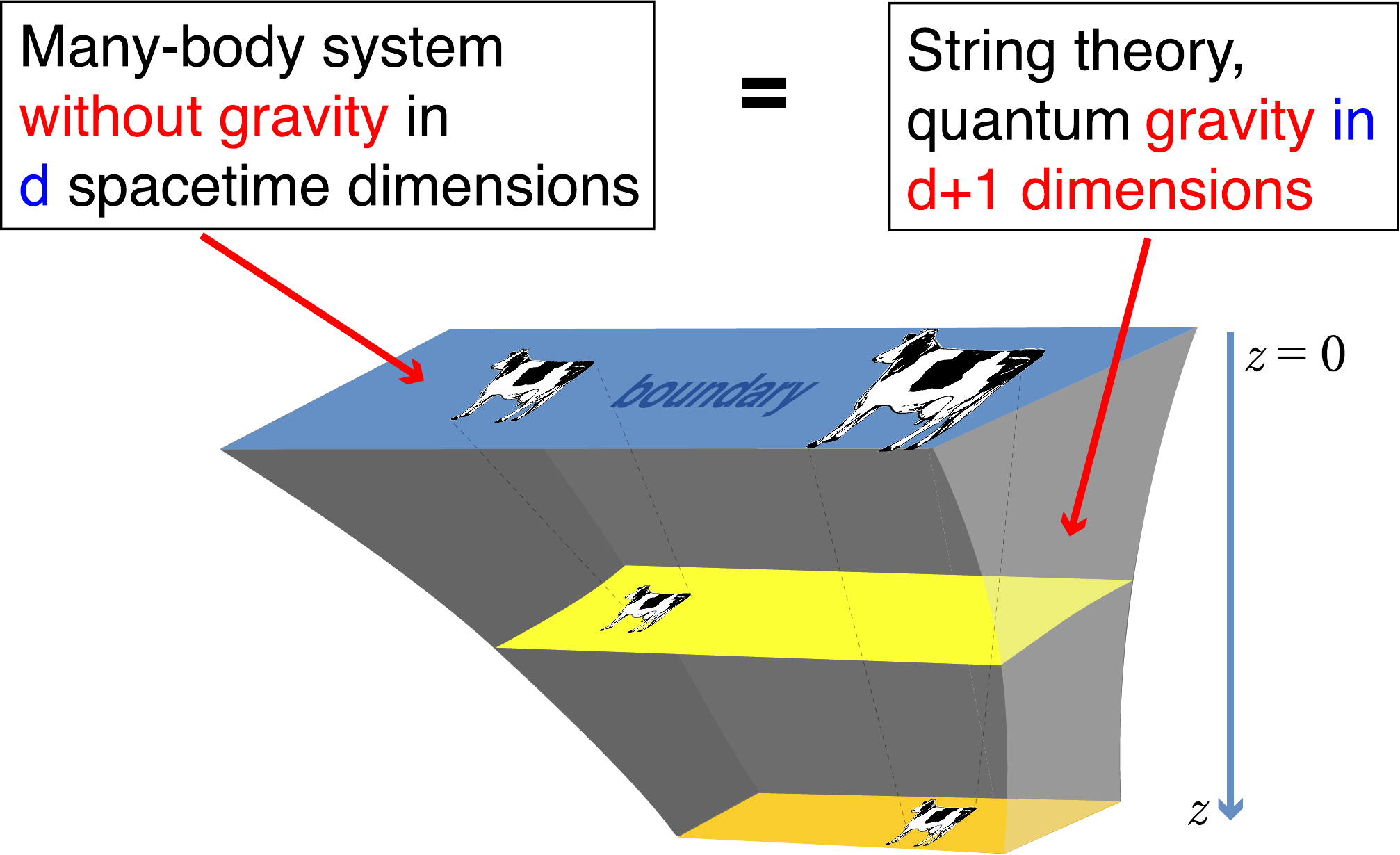}\\
\end{center}
\caption{In holographic duality, a quantum gravity system defined in a
$(d+1)$-dimensional anti-de Sitter spacetime is equivalent to a
many-body system defined on its $d$-dimensional
boundary.  Anti-de Sitter spacetime is a curved spacetime of
constant negative curvature. It has a radial
direction $z$ which runs from $0$ to $+\infty$, with a
$d$-dimensional Minkowski spacetime at each constant value of
$z$. $z = 0$ is the boundary of the whole spacetime (often referred to as the ``bulk'' spacetime) and is
where the many-body system is defined.  At a heuristic level, the radial direction $z$ in the bulk can be interpreted as corresponding to the size of structures in the boundary many-body system. 
For example, two objects in the bulk that are identical except for their
radial coordinate $z$ correspond in the boundary system to
two objects that are identical in all respects except for their size--one can be obtained from the other by magnification. 
This correspondence, with larger structures on the boundary corresponding to deeper structures in the bulk, 
is the key to how the boundary system can describe all the physics within the bulk even though it has one dimension less. 
By analogy, the boundary system is referred to as a ``hologram'' of the bulk system (since in laser physics a hologram is a two-dimensional representation of a three-dimensional object) and we say that there is a ``holographic duality'' between the boundary many-body system and the bulk quantum gravity system.
}
 \label{fig:duality}
\end{figure}

\begin{figure}[t]
\begin{center}
\includegraphics[width=.9\columnwidth]{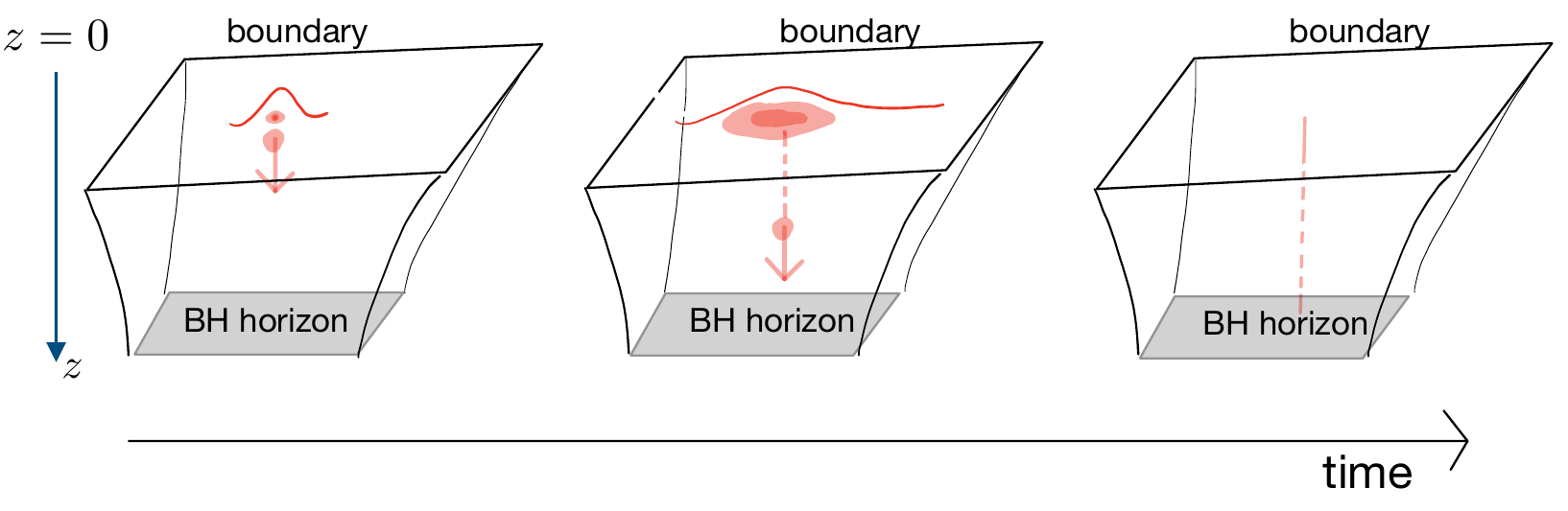}\\
\end{center}
\caption{Black holes and lack of quasiparticles.  Local disturbances applied to the boundary systems correspond to excitations of the gravity systems near the boundary. Once generated such excitations rapidly fall towards and are absorbed into the black-hole horizon. From the point of view of the boundary system this means that the disturbances are rapidly dissolved into the  `quantum soup' surrounding them, i.e. 
 before they are able to propagate to any appreciable distances, they have already dissipated. Technically it means that the width of any excitation is so large that it cannot be treated as a long-lived quasiparticle. 
}
 \label{fig:NoQuasiParticle}
\end{figure}

In holographic systems, i.e. quantum many-body systems with a gravity dual (see Fig.~\ref{fig:duality} for a brief description of the holographic duality), the absence of quasiparticle excitations is a general feature and transports in such systems generically exhibit the Planckian dissipation.
There is a simple geometric reason behind this effect. Holographic systems at a finite density/temperature are described by black holes, a defining feature of which is that they swallow everything thrown at them as fast as possible. Without fine tuning initial conditions such absorption processes happen with a time scale controlled by the geometric scale of a black hole which in turn determines its temperature. As a result the dissipation time is generically of the same order as the inverse temperature scale. See Fig.~\ref{fig:NoQuasiParticle}. Black holes thus provide a remarkably powerful description of certain strongly correlated systems without quasiparticles, for which many thermodynamic, transport, quantum informational, and far-from-equilibrium dynamical properties can be obtained in great detail.

In this paper we give a broad-stroke overview of recent developments of connections among gravity, strongly correlated many-body systems, and quantum information. 
We will mainly focus on the conceptual lessons learned and their intuitive origins, rather than technical details. 
The reader may also want to consult~\cite{Adams:2012th,DeWolfe:2013cua,Erdmenger:2018xqz,Liu:2018crr} for other recent reviews as well as the books~\cite{CasalderreySolana:2011us,nuastase2015introduction,ammon2015gauge,zaanen2015holographic,hartnoll2018holographic}.

After a brief description of essential elements of holographic duality in Sec.~\ref{sec.EssentialElements}, 
we discuss in Sec.~\ref{sec.StronglyCorrelated} the insights into phases and transports of strongly 
correlated systems from holography, in particular, the prediction of a universal intermediate energy phase from 
holography and a recent construction of Mott insulator.
 In Sec.~\ref{sec.scrambling} we discuss progress in understanding quantum many-body chaos and scrambling of quantum information from black holes. We will see that black holes are  ``fastest scramblers,''  in the sense that they are the most efficient systems to scramble quantum information~\cite{Hayden:2007cs,Sekino:2008he}. We will also discuss some features of maximally chaotic systems including the phenomenon of pole-skipping, and implications of chaos for energy transport. Finally we discuss 
 a phenomenon called ``regenesis'' resulting from a nontrivial combination of quantum chaos and entanglement, which 
 was originally discovered in gravity from the construction of a traversable wormhole. 

Given the geometric perspective offered by holographic duality, there has also been significant progress in understanding connections between spacetime structure and quantum information of holographic systems, which will be discussed in Sec.~\ref{sec:QI}.  
We will describe how to calculate fine-grained entanglement entropies using semi-classical gravity,  
 a ``local version" of the duality called subregion duality, and an understanding of quantum states with a semi-classical gravity description in terms of quantum error correcting codes. 
As an illustration of the power of these new insights, we discuss how they can be exploited to shed new light on the black hole information loss paradox.

Finally, in Sec.~\ref{sec:outlook} we conclude with a brief outlook.

\section{Essential elements of holographic duality}
\label{sec.EssentialElements}
The basic idea and the origin of the name ``holography'' are illustrated in Fig.~\ref{fig:duality} and its captions. 
Holographic duality was first discovered in the context of string theory in some special examples~\cite{Maldacena:1997re,Gubser:1998bc,Witten:1998qj}, but the general structure which has emerged from its study suggests that any quantum gravity theory in an asymptotic AdS spacetime should be described by some quantum many-body system~(often a conformal field theory). There are various types of examples of holographic dualities, all involving boundary systems with 
a large $\sN$ limit, where $\sN$ characterizes the number of degrees of freedom in the boundary system. 
The most well understood, involve systems whose degrees of freedom can be arranged in terms of $N \times N$ matrices (with thus $\sN \propto N^2$), such as non-Abelian gauge theories. 
A key feature of the duality for such ``matrix-type'' systems is that there exists a strongly coupled regime which corresponds on the gravity side to the regime of classical Einstein gravity plus various matter fields. 
This makes it possible to extract in great detail many physical properties of the boundary systems using semi-classical gravity analysis.  

There are  other types of duality examples. One type involves boundary systems being vector-like theories, e.g. $O(N)$ vector models, which are believed to be dual to a bulk theory with an infinite number of massless higher spin fields~\cite{Klebanov:2002ja,Giombi:2009wh,Vasiliev:2003ev}. Another type is the Sachdev-Ye-Kitaev model~\cite{Sachdev:1992fk,Kitaev-talks:2015}, which consists of $N$ Majorana fermions interacting via random couplings of zero mean, which is believed to be dual to a two-dimensional gravity theory coupled to an infinite tower of fields~\cite{Polchinski:2016xgd,Maldacena:2016hyu,Maldacena:2016upp,Kitaev:2017awl,Engelsoy:2016xyb,Jensen:2016pah,Jevicki:2016bwu,Jevicki:2016ito}.
In neither of these types of boundary systems, the corresponding bulk dual has a regime which reduces to 
the Einstein gravity.

In this review we will be mostly concerned with ``matrix-type'' systems whose dual has an Einstein gravity regime.  In this section we first highlight some pieces of the dictionary between a dual pair, and then elaborate on some key features.
We will take the boundary spacetime dimension to be $d$, without restricting to a specific dimension.

\subsection{Basic dictionary}

The classical action for the bulk gravitational theory can be written as 
\be \label{nmb}
S_{\rm bulk} =  S_{\rm grav} + S_{\rm matter}
\ee
with $S_{\rm matter}$ the action for possible matter fields, and $S_{\rm grav}$ the gravitational action with a negative cosmological constant 
\be\label{eq.EHaction}
S_{\rm grav} ={1 \ov 16 \pi G_N}  \int d^{d+1}x \sqrt{-g} \left( R + \frac{d(d-1)}{\ell^2} \right)\ .
\ee 
In~\eqref{eq.EHaction}, $\ell$ is a length scale and Newton's 
constant $G_N$ is inversely proportional to the number of degrees of freedom $\sN$ of the 
boundary theory. Thus gravity is weak if $\sN$ is large. 
The  spectrum of matter fields and the specific form of $S_{\rm matter}$ vary with the 
specific dual boundary system, while the gravity action $S_{\rm grav}$ is universal to all systems with an Einstein gravity dual.

The gravity discussion will be restricted to the semi-classical limit, i.e. treating the gravitational theory perturbative in powers of $\hbar G_N \to 0$ with the leading order being the classical gravity. On the boundary system, this corresponds to expanding in 
$1/\sN$ with $\sN \to \infty$. 

The classical bulk action~\eqref{nmb} is also corrected by higher derivative terms, suppressed by inverse powers of the Planck mass or, if present, further dimensionful parameters, such as the string tension in many well-established examples of dual pairs.
It is  often important to understand whether or not certain qualitative physical phenomena deduced in holography are stable under such corrections, albeit perhaps with potentially small corrections to numerical values of physical parameters. We shall have occasion to comment on such potential corrections in later sections.

The equivalence of Fig.~\ref{fig:duality} means that there should be a one-to-one correspondence between quantum states
between the bulk and boundary systems.  On the gravity side, states are represented in the classical limit by solutions to equations of motion of~\eqref{nmb}  with appropriate boundary conditions, each of which thus should correspond to some quantum state of the boundary system. As an illustration let us look at some simple solutions to~\eqref{nmb} with no matter excited~(which then reduce to solutions of~\eqref{eq.EHaction}). 
The boundary descriptions of these solutions are independent of specific systems and thus are universal among all systems with a gravity dual. 

The simplest and most symmetric solution is the anti-de Sitter spacetime, 
\be\label{eq.poincareadsmetric}
ds^2 = \frac{\ell^2}{z^2} \left( -dt^2 + dz^2 + \sum_{i=1}^{d-1}dx_i^2 \right) \ ,
\ee
which corresponds to the {\it vacuum} of a dual conformal field theory.
Below we will use the notations $x^M  = (z, x^\mu)$ where $x^{\mu} = (t, x_i)$ run over the coordinates of the boundary theory. $z \in (0, +\infty)$ is the extra ``holographic'' coordinate, with the AdS boundary located at $z=0$ (as $z \to 0$, the overall prefactor $1/z^2$ in~\eqref{eq.poincareadsmetric} blows up analogous to $r \to \infty$ limit of 
a flat Euclidean metric $ds^2 = r^2 d \Om^2$ in spherical coordinates), while large values of $z$ can be considered as the ``interior'' of AdS (see Fig. \ref{fig:duality}).  The metric~\eqref{eq.poincareadsmetric} has a large number of isometries (i.e. coordinate transformations which leave the metric invariant), which are in one-to-one correspondence with conformal transformations of  
the boundary system. Among all the isometries of~\eqref{eq.poincareadsmetric} we would like to draw particular attention to the following scaling symmetry
\be
z \to \lambda z \qquad x^{\mu} \to \lambda x^{\mu} \ . \label{scalisom}
\ee
From~\eqref{scalisom} as we scale the boundary coordinates $x^\mu$ we must accordingly scale the radial coordinate $z$. This indicates that $z$ represents length scales of the boundary theory: we scale to short distances (UV) in $x^\mu$ as $z$ scales to $0$, and to long distances (IR) in $x^\mu$ as $z$ scales to $\infty$. In other words, going from the boundary $z=0$ to larger values of $z$ along the radial direction may be considered as going from UV to IR in the boundary system. This turns out to be a general feature of all bulk geometries, including those for which~\eqref{scalisom} is no longer an isometry. 
Recall that the central idea of the renormalization group (RG) is to organize the physics of a many-body system in terms of scales, thus 
the radial direction of AdS can be considered as a geometrization of renormalization group (RG) flow of the boundary theory!

Another simple solution to~\eqref{eq.EHaction} is the Schwarzschild black hole
 \be \label{equi1}
 ds^2 = \frac{\ell^2}{z^2} \left( -f(z) dt^2 + \frac{dz^2}{f(z)} + \sum_{i=1}^{d-1} dx_i^2  \right)\,,
 \ee
 where $f(z) = 1 - \left({  z \ov z_h}\right)^d$ and $z_h$ is a constant. Equation~\eqref{equi1} has an event horizon at $z = z_h$ with  topology 
 $\RR^{d-1}$.  See  Fig.~\ref{fig:NoQuasiParticle} for a cartoon of the black hole geometry. 
 From the discoveries of Hawking and Bekenstein in the 1970s,  black holes are known to be thermodynamic objects, which makes it natural to identify~\eqref{equi1} with a thermal state of the boundary system,  
 with its Hawking temperature $T_H = {d \ov 4 \pi  z_h}$  identified with the boundary system temperature. 
 Note that as $z \to 0$, $ f(z) \to 1$, equation~\eqref{equi1} reduces to~\eqref{eq.poincareadsmetric}. This is consistent with the above discussion of $z$ as representing length scales: as $z \to 0$ we go to short distances and recover vacuum physics, while 
 the whole geometry~\eqref{equi1} tells us how the system flows from vacuum physics at short-distances (UV) to thermal physics at IR scales. The presence of an event horizon at some finite value of $z=z_h$ can be considered as an IR ``cutoff'' representing the inverse temperature scale. In contrast, in~\eqref{eq.poincareadsmetric}  the values of 
 $z$ extend all the way to $+\infty$, reflecting that near the vacuum, there exist excitations of arbitrarily low energies.
 
A second aspect of the dictionary is the correspondence between ``elementary'' bulk fields appearing in~\eqref{nmb} and boundary operators. 
Since different boundary theories have different operator spectra,  the precise dictionary depends on 
specific systems. Nevertheless,  there are some common elements universal to all theories: (i) 
the boundary stress tensor $T^{\mu \nu}$ is dual to the spacetime metric  $ g_{MN}$; (ii) 
 a conserved boundary current $J^\mu$ is dual to a bulk 
gauge field $A_M$ whose action is given by the bulk Maxwell or Yang-Mills action. 
Using the field/operator correspondence one can calculate, say,  
correlation functions of an operator $\sO$ in the strongly coupled boundary systems from weakly coupled dynamics (since $G_N$ is small) of its dual field $\phi$ on the gravity side. In particular, the transport coefficients associated with conserved quantities $T^{\mu \nu}$ and $J^\mu$ are obtained from the dynamics of $g_{MN}$ and $A_M$ which often exhibit universal behavior for all holographic systems with a gravity dual.

\begin{figure}[t]
\begin{center}
\includegraphics[width=.9\columnwidth]{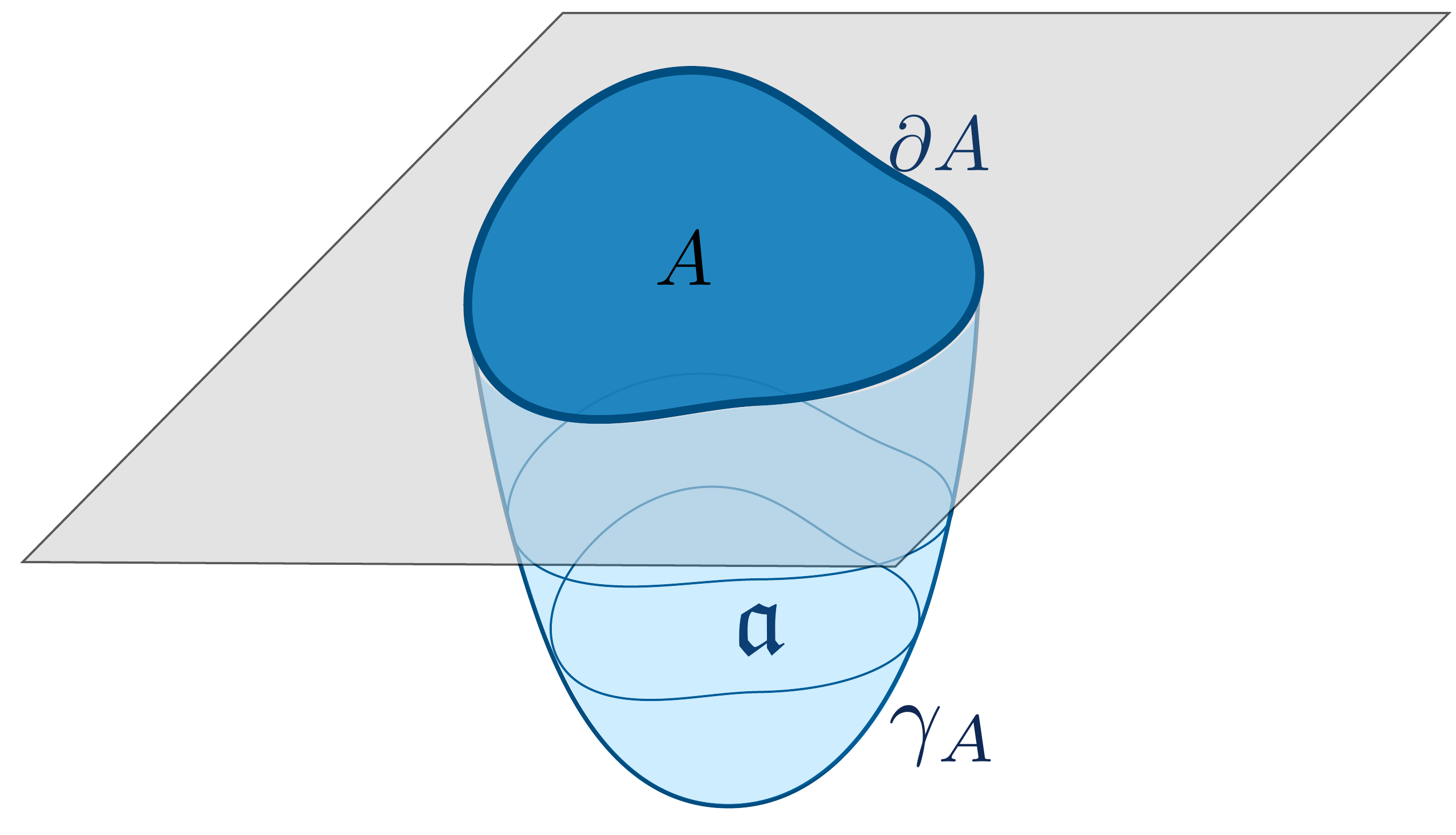}\\
\end{center}
\caption{Minimal surface $\ga_A$ for the the Ryu-Takayanagi formula~\eqref{eq.RTformula}. 
The figure shows a single time slice of the bulk spacetime. 
The spatial region $\fa$ between $\ga_A$ 
and $A$, i.e. $\p \fa = \ga_A \cup A$,  is often referred to as the ``entanglement wedge'' associated with $A$ and will play an important role in Sec.~\ref{sec:QI}. (More precisely, the entanglement wedge refers to the domain of dependence of $\fa$, but here for language simplicity we will not make this distinction.)
}
 \label{fig:RTEntanglementWedge}
\end{figure}

A third piece of the holographic dictionary we would like to highlight is concerned with how quantum information of a boundary system in encoded in the gravity theory.  Suppose the boundary system is in a state described by some density operator $\rho$ and consider a spatial subregion $A$.   Imagine a UV regularization (say putting the system on a lattice) such that the Hilbert space factorizes into ${\cal H} = {\cal H}_A \otimes {\cal H}_{\overline A}$, where $\overline A$ denotes the complement of $A$.  The entanglement entropy $S_A$ of subregion $A$, is then defined as 
 the von-Neumann entropy of the reduced density matrix $\rho_A = {\rm Tr}_{\bar A} \rho$, i.e. 
 $S_A = S(\rho_A) =  -{\rm Tr}_A \rho_A\log\rho_A$.  
In a many-body system (including non-interacting systems!), computing $S_A$ is in general a very difficult task. One typically proceeds by computing first the R\`enyi entropy $S_n = \frac{1}{1-n}{\rm log}\, {\rm Tr}_A \rho_A^n$  for $n\in \mathbb{Z}$ and then using the so-called replica trick~\cite{Holzhey:1994we,Calabrese:2004eu}, whereby one takes the limit $n\rightarrow 1$ formally by analytically continuing the result away from integer values. In holographic duality,  $S_A$ can be directly obtained without using the replica trick, in terms of a beautiful geometric formula first proposed by Ryu and Takayanagi~\cite{Ryu:2006bv} 
\be\label{eq.RTformula}
S (\rho_A) = \frac{{\rm Area}(\gamma_A)}{4\hbar G_N}\, . 
\ee
For a static state $\rho$, $\ga_A$ in the above equation is a spacelike codimension-2 surface in the bulk geometry dual to state $\rho$  satisfying the following conditions: (i) its boundary coincides with boundary of $A$, i.e. $\p \ga_A = \p A$ (where $\p A$ denotes the boundary of $A$); (ii) it is homologous to $A$, i.e. $\ga_A \cup A = \p \fa$ where $\fa$ is some bulk region; (iii) 
it lies on the same time slice as the boundary subregion $A$; (iv) it has the smallest area among all surfaces satisfying (i)--(iii), i.e. a minimal surface.  See Fig.~\ref{fig:RTEntanglementWedge}.
For a general time-dependent state or the subregion $A$ not lying in a static time slice, one should replace conditions (iii) and (iv) above by the condition that the area should be extremized among all surfaces satisfying (i) and (ii)~\cite{Hubeny:2007xt}. 
When there are multiple extremal surfaces, one picks the one with the smallest area.

It is striking that, entanglement entropy, which is a fine-grained quantum informational measure,
can be captured by a classical geometric quantity of the corresponding bulk solution, a coarse-grained measure, without the need of knowing any microscopic details of the bulk gravity theory. Equation~\eqref{eq.RTformula} thus provides a ``hydrodynamic'' description of entanglement via spacetime geometry. Conversely, the spacetime structure may be considered as a carrier of quantum information of the underlying quantum system, i.e. the fabric of spacetime is quantum information. 
The RT formula~\eqref{eq.RTformula}  provides a starting point 
for many further relations between gravity and quantum information to which we will return in Sec.~\ref{sec:QI}.

\subsection{Anatomy of boundary systems by scales}

In addition to providing new powerful techniques for calculating observables in certain strongly correlated systems, holography provides a higher dimensional ``panoramic view'' 
of a quantum many-body system, 
with physics at all scales presented in parallel.  
A striking feature of such a ``panorama'' is that 
many highly nontrivial quantum dynamical 
aspects of the boundary system are described on the gravity side in terms of simple geometric pictures. 
As a result, the duality gives rise to new perspectives and new probes of the boundary system, which are not possible using conventional methods, even if one could solve the system exactly.

\begin{figure}
\begin{center}
\includegraphics[width=\columnwidth]{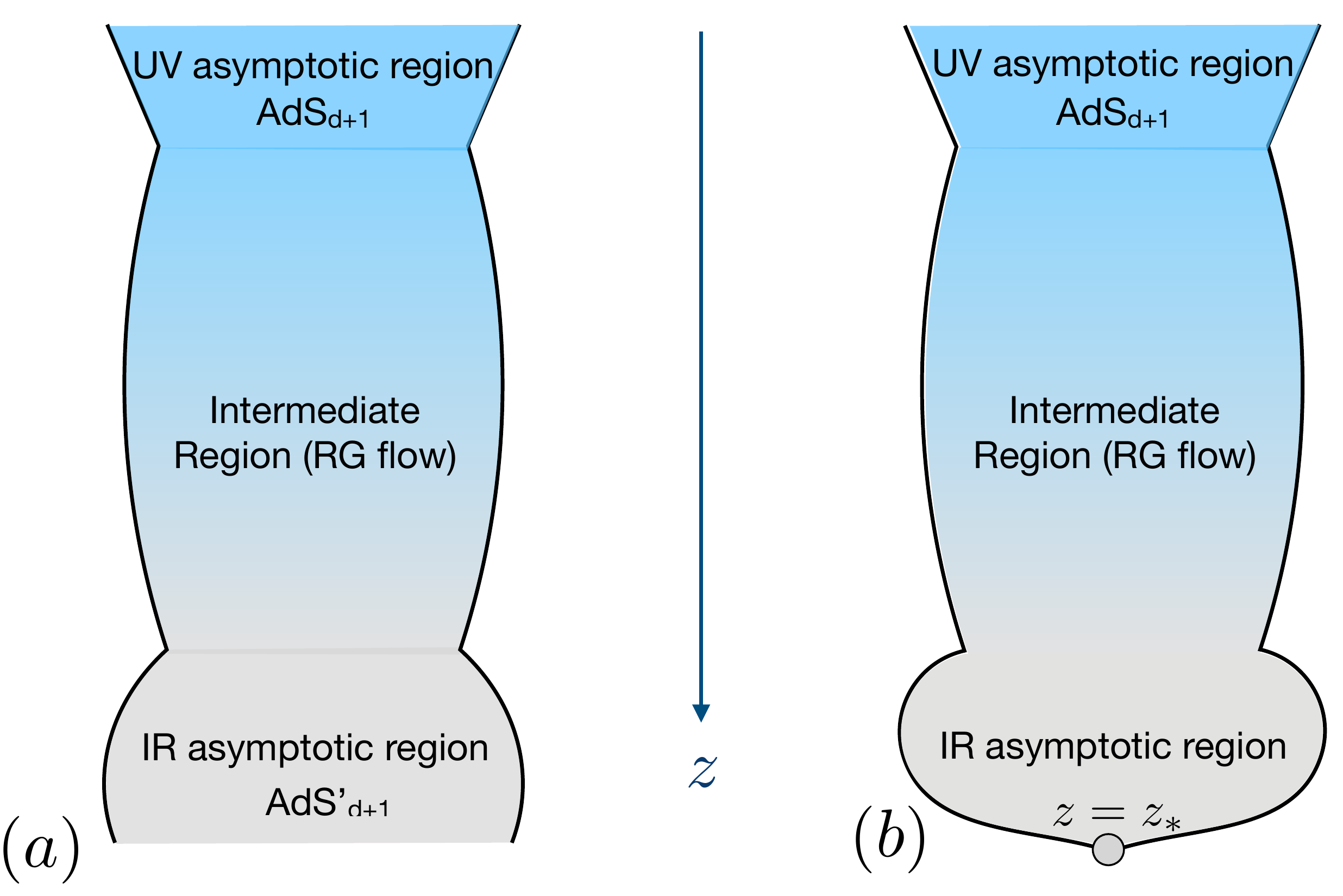}\\
\end{center}
\caption{Geometric features of gapped and gapless systems.
(a): the geometry represents the flowing from a UV fixed point to an IR fixed point. 
The geometry becomes scale invariant and is described by~\eqref{eq.poincareadsmetric} as $z \to 0$ and $z \to \infty$ with different curvature radii $\ell$. The intermediate region between them describes the flow between the two fixed points. (b): the geometry represents an RG flow where the UV fixed point flows to gapped theory in the IR, ending smoothly at a maximum radius $z_*$. Another option is that $z_*$ is a singularity, which then needs to be resolved in order to understand the nature of the ground state.}  
 \label{fig:AdS}
\end{figure}

As a simple illustration, in Fig.~\ref{fig:AdS} (a) and (b) we present a contrast of geometric differences  
between the gravity descriptions of the vacuum state of a gapless system and a gapped system.
In a quantum many-body system, determining whether the vacuum is gapless or gapped is a highly 
nontrivial dynamical question, but we see that it is reflected in a very simple geometric feature of the bulk 
geometry. Furthermore, the bulk geometry represents the full ``evolution'' of the boundary system from UV to IR scales 
through the ``evolution'' of its geometry along the radial direction.
Life is, however, not always so effortless, going to the interior of AdS, we often encounter a time-like curvature singularity as discussed in the caption of Fig.~\ref{fig:AdS}. The nature of the ground state of the boundary then depends on how the singularity is resolved. Resolving curvature singularities of classical solutions in general relativity is one of the most important goals of quantum gravity. Thus in such a case a highly nontrivial dynamical IR problem in the boundary system is mapped 
to a dynamical UV problem in quantum gravity. This presents new exciting opportunities on both sides. In the case the boundary theory result is known one can then gain important hints on how a singularity is resolved in gravity (see e.g.~\cite{Klebanov:2000hb}). 

The description of entanglement entropies of a boundary system~\eqref{eq.RTformula} is another example of how highly quantum dynamical information 
of the boundary system is encoded in simple geometric quantities of the bulk theory.

\section{Finite density phases and charge transports}\label{sec.StronglyCorrelated}

In this section we give a brief overview of different phases which can arise in holographic systems at a finite charge density
and charge transports.

\subsection{The zoo of holographic phases}

To pursue connections  to condensed matter physics, we consider holographic systems with a finite charge density, 
which can be achieved by taking a system with a $U(1)$ global symmetry and turning on a finite chemical potential $\mu$ for the conserved charge. 
For a general strongly interacting many-body system, understanding the IR physics resulting from such a (relevant) deformation is a very difficult question, and indeed lies at the heart of the challenges for treating strongly correlated electronic systems. 
For holographic systems, it turns out there is a simple geometric description 
in terms of a {\it charged} black hole whose metric also has the form~\eqref{equi1}, but with a modified function $f(z)$. The Hawking temperature of the black hole is again identified with the boundary temperature $T$ and 
the presence of the chemical potential $\mu$ is reflected in the black hole now carrying an electric charge under the Maxwell field dual to the conserved current. The charged black hole geometry is homogeneous and isotropic, and is a universal description of a finite-density state independent of specific boundary systems.

To extract the IR physics of the boundary system, one should examine the near-horizon region of the charged black hole geometry in the low temperature regime  $T/\mu \ll 1$. Interestingly, the near horizon geometry is given by AdS$_2 \times \RR^{d-1}$ (below $\ell_2$ is the curvature radius for AdS$_2$ and $b$ is a constant)
\be \label{ads2}
ds^2 =  \frac{\ell_2^2}{\zeta^2}(-dt^2 + d\zeta^2) + b^2 \sum_{i=1}^{d-1} dx_i^2 
\ee
 which indicates that the system flows to a nontrivial IR fixed point dual to gravity in~\eqref{ads2}. The IR fixed point, which has been referred to as a semi-local quantum liquid (SLQL)~\cite{Iqbal:2011in}, can be characterized by the following features: (1)~it is invariant under a scaling $t \to \lam \, t $ only in time ($\lam$ an arbitrary constant) with spatial coordinates $\vec x$ unchanged. (2)~It has a finite correlation length $\xi \propto \mu^{-1}$. 
(3)~One can classify observables by their IR scaling dimension $\De$ under a time scaling. $\De$ depends nontrivially on $k/\mu$ where $k$ is the magnitude of the spatial momentum of an operator (this is allowed since spatial directions do not scale). 
(4)~The system is highly entangled, with the entanglement entropy of a subregion proportional to its volume rather than 
the area in the low temperature limit. (5)~There is a residual thermodynamic entropy in the zero-temperature limit. 

The above features of SLQL lead to many phenomenological properties (see~\cite{Iqbal:2011ae} for a technical review) parallel to those of the strange metal phases of high-temperature superconducting cuprates and 
heavy fermion compounds, with the time scaling reminiscent of the ``local quantum criticality'' observed in these materials~\cite{varma1989phenomenology,coleman99,Si.01,mitrano2018anomalous}. 
For example, studies of fermionic excitations in SLQL reveal novel non-Fermi liquids without quasi-particles~\cite{Liu:2009dm,Cubrovic:2009ye,Faulkner:2009wj}. Furthermore, for fermonic operators of certain IR dimension, one recovers~\cite{Faulkner:2009wj} the Marginal Fermi liquids description~\cite{varma1989phenomenology} for cuprates and a linear resistivity~\cite{Faulkner:2010zz,Faulkner:2013bna}. Studies of bosonic instabilities 
in SLQL lead to a variety of other phases, and patterns of quantum phase transitions similar to various heavy fermion compounds~\cite{Faulkner:2010gj,Iqbal:2011aj,Jensen:2011af}.

In fact, essentially all known finite density phases in holographic systems can be viewed as being originated from instabilities of SLQL
one way or the other. When an instability occurs for a charged scalar, one obtains a superfluid or a superconductor~\cite{Gubser:2008px,Hartnoll:2008vx}, and for neutral scalars  phases resembling an anti-ferromagnet~\cite{Iqbal:2010eh} or metamagnetic phases~\cite{Donos:2012yu,DHoker:2010zpp}. It can also develop certain ``fermionic stabilities'' to transition to a Fermi-liquid phase~\cite{Iqbal:2011in,Hartnoll:2011dm}. 

Instabilities can also happen to a (charged or neutral) scalar at a non-zero wavevector, which leads to a variety of 
spatially modulated superfluids or charge density wave like phases, see e.g.~\cite{Nakamura:2009tf,Ooguri:2010kt,Donos:2011bh,Donos:2011ff,Rozali:2012es,Rozali:2013ama,Donos:2013wia,Withers:2013loa,Withers:2014sja,Bu:2012mq,Donos:2015eew,Cai:2017qdz}. 
 It is also possible to have insulators~\cite{Donos:2012js,Andrade:2017ghg} and phases which maintain 
homogeneity while breaking translations~\cite{Donos:2012wi}.
 
 Thus  holographic matter at a finite density exhibits a fascinating wealth of ordered phases and spatially modulated states, with the SLQL as their ``mother'' phase. Holographic duality 
maps the problem of classifying possible phases to classifying solutions to the Einstein equations coupled to various matter fields, providing a powerful lens into the landscape of strongly correlated states of matter, much of which still remains to be explored (see~\cite{Iizuka:2012iv} for attempts at a classification). 

That the SLQL has a residual thermodynamic entropy in the zero-temperature limit suggests that 
it cannot be a genuine ground state of the boundary system at a finite chemical potential. Indeed there are arguments on both 
the gravity~\cite{Preskill:1991tb} and boundary~\cite{Jensen:2011su} sides that such a state cannot be extended to 
sufficiently low temperatures. Accordingly one should view SLQL as an intermediate energy phase, controlling physics for a range of temperatures $T\ll \mu$, but not $T=0$. 
 Its residual thermodynamic entropy in the $T \to 0$ limit should be viewed as an artifact of the large $\sN$ approximation one is using in the classical gravity regime and should be understood as reflecting a set of 
closely spaced intermediate-energy states above the genuine ground state.

\begin{figure}[h]
\begin{center}
\includegraphics[scale=0.3]{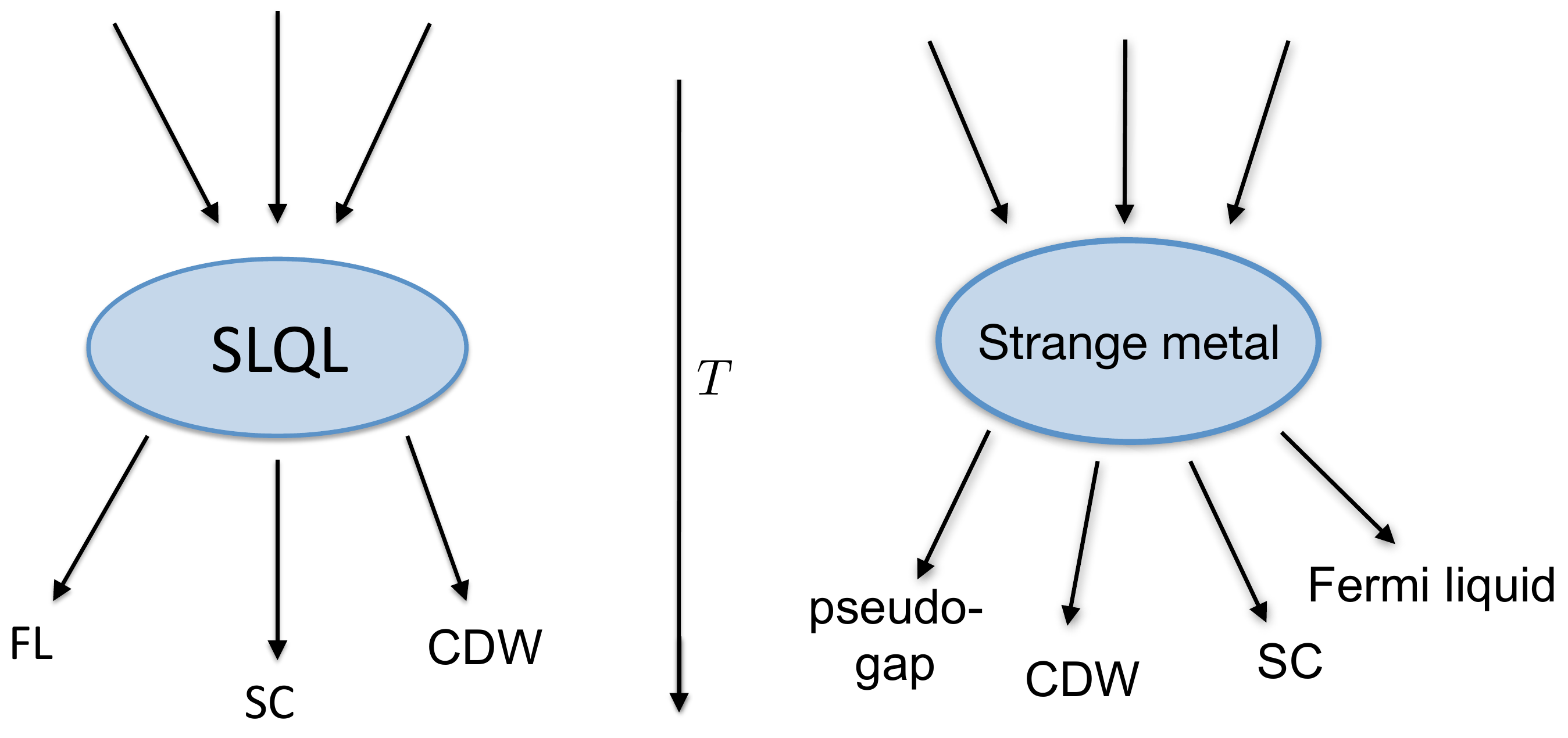}
\end{center}
\caption{{\bf Left}: The semi-local quantum liquid phase as a universal intermediate energy phase. Many different UV systems flow to it, and at lower temperatures depending on parameters it settles into one of many different possible ground states, such as Fermi liquids (FL), anti-ferromagnets (AFM), and superconductors (SC) among others. This is reminiscent of the phase diagram for hole-doped cuprates, where a cartoon is shown on the {\bf Right}: materials of different microscopic structures flow to the same strange metal phase and then depending on parameters such as doping go over to a variety of other phases at lower temperatures. 
}
\label{fig:phase}
\end{figure}

It has been argued in~\cite{Iqbal:2011in} that the SLQL appears as a mother phase for many other holographic phases is not an accident, but should be viewed as a prediction from holography for the existence of {\it universal intermediate energy phase}, in the sense that systems with very different microscopic (UV) physics and very different ground state structures all go through SLQL in some intermediate range of temperatures. See Fig.~\ref{fig:phase}, such a picture is also strongly reminiscent of the strange metal phases for the cuprates and heavy fermion compounds and suggests that a universal intermediate energy phase like SLQL may underlie the strange metal phases for these real-life materials. It is also tempting to speculate that being highly entangled could lie behind the potential for SLQL to ``carry'' a large variety of cohesive ordered states.

\iffalse 
In the AdS$_2$ case, the mechanism of these instabilities proceeds intuitively as follows: AdS itself is able to support fields whose mass is slightly tachyonic, so long as its square remains above a lower bound first determined by Breitenlohner and Freedman \cite{Breitenlohner:1982jf}. But it may happen that somewhere down the IR  AdS$_2$throat geometry, a field which would be stable in the UV develops an effective mass that falls below the local (IR) BF bound.  To give but one example, one can show explicitly that this is the mechanism of the first holographic ordered phases to have been discovered, namely the holographic superconductor \cite{Gubser:2008px,Hartnoll:2008vx}. \js{instabilities of IR scaling geometries?}
\fi 

Recently there has been exciting progress in constructing phenomenological Hamiltonians based on the SYK models~\cite{Sachdev:1992fk,Kitaev-talks:2015,Sachdev:2015efa} to describe the strange metal phase of cuprate materials~\cite{Song:2017pfw,Ben-Zion:2017tor,Patel:2017mjv,Chowdhury:2018sho,Altland:2019lne,Altland:2019fyj,Blake:2017qgd}. These models 
share striking similarities with the phenomenology of SLQL discussed above and may be considered as explicit
Hamiltonian constructions of such a universal intermediate energy phase. The similarities between these SYK-motivated models and the holographic strange metal is not an accident. It has been argued previously in~\cite{Sachdev:2010um,Sachdev:2015efa} that the SYK model is in fact holographically dual to gravity in AdS$_2$ spacetime which 
is the key component of the SLQL physics. A difference among holographic systems and these models has to do with how to ``weave'' the SYK quantum dots into higher dimensional quantum systems.

In addition to AdS$_2 \times \RR^{d-1}$, which gives to the SLQL, 
holographic systems can also have other scaling intermediate phases with a general 
dynamical exponent. With dynamical exponent $\zeta$ defined in terms of scaling  
\be
t \to \lambda^\zeta t,\qquad x^{i} \to \lambda x^{i}  \label{eq.LifshitzScaling}
\ee
the SLQL may be interpreted as having $\zeta = \infty$. By turning on the expectation value of a marginal scalar operator 
in addition to the chemical potential, one could obtain black holes whose near-horizon geometries are parameterized by two other independent parameters $\zeta$ (the dynamical exponent) and $\theta$ (hyperscaling violating parameter)~\cite{Gubser:2009qt,Goldstein:2009cv,Goldstein:2010aw,Charmousis:2010zz}. In such systems, the entropy scales with temperature as 
$S\sim T^{d  -\theta \ov \zeta}$ instead of $S\sim T^{d \ov \zeta}$ for a system invariant under scaling~\eqref{eq.LifshitzScaling}
and thus the name  hyperscaling violating parameter for $\theta$~\cite{Huijse:2011ef}. 
This family of solutions should again be interpreted as describing some intermediate-energy phases
as they all have time-like singularities in the interior. The precise ground state structures depend on the resolution 
of the singularities. Applying the null energy condition (i.e. energy in classical gravity cannot be negative) to these solutions 
results in the interesting constraints on $\zeta$ and $\theta$
$(d-\theta) \left(d(z-1) - \theta \right) \ge 0$ and 
$(z-1) \left( d+z - \theta \right)\ge 0$~\cite{Dong:2012se},
whose precise physical interpretation remains somewhat mysterious from the boundary perspective.

\subsection{Charge transport}

A finite charge density system with translation symmetry generally has an infinite DC conductivity 
as there is in general a nontrivial overlap between the current operator and momentum operator, which is conserved. 
The same thing happens to systems where translations are broken only spontaneously, 
as there remains a sliding mode corresponding to a translation of the whole system as a rigid structure (see e.g.~\cite{Donos:2018kkm} for examples in holographic systems). A finite conductivity can result in a translation invariant situation 
if the system is neutral (i.e. particle-hole symmetric) in which case the current operator has no overlap with momentum, or 
if the charged system is coupled to an infinite bath through which the momentum can be dissipated. 

Thus to have a finite conductivity for a full charged system (i.e. without bath) one needs to break momentum conservation 
explicitly. This can be achieved through introducing random disorder, or putting the system on a lattice, or some effective
macroscopic description which breaks momentum conservation. Introducing random disorder in the gravity 
description of a boundary system is intricate \cite{Adams:2011rj,Adams:2012yi,Hartnoll:2008hs,Lucas:2014zea,Lucas:2015lna} and usually requires tremendous numerical power~(see e.g.~\cite{Hartnoll:2014cua}).  There is a beautiful way to introduce momentum non-conservation in a translation invariant 
setting by working with an effective description of the gravity theory in which graviton has a phenomenological mass~\cite{Vegh:2013sk} (see also~\cite{Davison:2013jba}). Such a ``massive gravity'' theory may be considered as an effective theory after one integrates 
out microscopic degrees of freedom responsible for momentum non-conservation which could be due to an external lattice or disorder. The results obtained in this approach are consistent with those obtained by introducing explicit lattices which we will discuss more explicitly below. 

Both from the perspective of simulating real-life condensed matter systems which have a lattice structure and 
having a finite conductivity, there has been much effort to put gravity systems on a lattice~\cite{Horowitz:2012ky,Erdmenger:2013zaa,Chesler:2013qla,Donos:2013eha,Alberte:2017oqx,Ling:2013aya,Donos:2014oha,Andrade:2015iyf,Andrade:2017leb}. One way to achieve 
this is to introduce a spatially varying chemical potential, e.g.  $\mu(x) \propto \cos(k x)$. As an illustration, we briefly describe the results in the so-called axion models, which are representative of the general structure one finds.
In this model translation symmetry is broken by spatial modulation of a scalar field (often called axion) and is  parametrized by a parameter $\alpha$ characterizing the strength of breaking. From gravity calculations one finds that the conductivity $\sigma_{\rm DC}$ separates into two distinct contributions~\cite{Andrade:2013gsa}, 
\bea
\sigma_{\rm DC} &=& \mu^{d-3} g\le(\frac{T}{\mu},\frac{\alpha}{\mu}  \ri)  \left(1 +  (d-2)^2 \frac{\mu^2}{\alpha^2} \right) \nonumber\\
&=&  \sigma_{\rm inc.} + \sigma_{\rm coh.}  \ .
\label{1o}
\eea
where $g(x,y)$ is a dimensionless function determined by the black-hole solution of \cite{Andrade:2013gsa}.
The first term above is independent of momentum dissipation and comes from {\it incoherent} production of particle-hole pairs, which is a consequence of the ``quantum critical'' nature of holographic systems. 
 The second term comes from momentum dissipation due to lattice effects~\cite{Davison:2014lua,Davison:2015bea}. 
When the breaking strength $\al$ is small, momentum dissipation is characterized by a long momentum dissipation time 
scale, $\tau_{\rm diss} \propto {1 \ov \al^2}$. The second term then dominates the conductivity and 
the optical conductivity $\sigma(\omega)$ exhibits a sharp Drude peak, despite the absence of quasiparticles~\cite{Hartnoll:2012rj}. 
In the opposite limit of large $\al$ (i.e. $\al \sim \mu$), the two terms in~\eqref{1o} are comparable and one can no longer separate the momentum dissipation time scale with the Planckian dissipation time $\hbar \beta$. Accordingly 
the corresponding optical conductivity exhibits a broad structure as in an incoherent metal (see e.g.~\cite{emery}).  It is interesting to note that the holographic conductivity~\eqref{1o} is a simple sum of contributions from two separate ``channels.'' This should be contrasted with the Matthiessen's rule for conventional metals 
which says that the total resistivity can be written as a sum of contributions from different electronic scattering processes such as with impurities and phonons, and has its origin in treating charge carriers in terms of quasiparticles. The ``sum rule''~\eqref{1o} highlights its non-quasiparticle origin.

A very interesting feature of holographic computations of transport coefficients such as conductivities is that
they can be deduced from universal horizon physics \cite{damour,membranebook}, due to simple flow equations that relate those horizon values to field theory transport coefficients \cite{Iqbal:2008by}. For certain transport properties these flow equations are actually trivial, so that the horizon value is precisely equal to the field theory value, a celebrated example being the ratio of shear viscosity over entropy density~\cite{Policastro:2001yc}, $\frac{\eta}{s} = \frac{1}{4\pi} = \frac{\eta_{\rm mb}}{ s_{\rm mb}}$ where the quantities $\eta_{\rm mb}$ and $s_{\rm mb}$ are the values of the corresponding transport coefficients of the effective `membrane' describing the horizon. A major advantage of such relations is that the horizon physics is much easier to understand than the full spacetime, and furthermore shows a great deal of universality, which is understood as a consequence of General Relativity itself \cite{damour,membranebook}.  Amazingly a similar physical picture exists for cases where translation invariance is broken, where again the problem of determining transport properties of the boundary field theory is reduced to a fluid/membrane-like picture at the horizon \cite{Donos:2014uba,Donos:2014cya,Lucas:2015lna,Banks:2015wha}.

\subsection{From metals to (Mott) insulators}

High temperature superconducting cuprates can be viewed as ``doped Mott insulators.'' 
Thus it is of great interests to construct Mott-insulator-like phases using gravity and examine what happens if one dopes 
them.  There has been exciting recent progress~\cite{Andrade:2017ghg} in this direction which we now briefly review. 

The understanding of conventional Mott insulators is based on the notion of a single charge carrier per unit cell.  
But in holographic systems any information regarding microscopic 
constituents has been completely washed out in the gravity description of 
any macroscopic states, for which therefore a more general formulation of a Mott insulator is needed. 
One can alternatively define a Mott insulator for a general quantum liquid
as a phase in which a charge lattice (determined from IR dynamics) is commensurate with an external 
background lattice. Clearly this contains the usual definition as a special case.  

To construct a Mott insulator in holography then requires that we set up the gravity system as follows: (1) The parameters 
of the system should be such that at sufficiently low temperature the system is in a charge density wave like phase (as discussed in our earlier treatment of instabilities of SLQL), with period $p_0$. We will refer to this periodic structure as the ``IR'' lattice.  (2) Impose an explicit background lattice with a fixed period $k$ in the same way as the ``lattice'' models for conductivities discussed above. We will refer to this external periodic potential the ``UV'' lattice. The state of the system is a consequence of 
the nontrivial dynamical interplay between the UV and IR lattices. The presence of UV lattice forces the period of the IR lattice to align with its period, i.e. deforming $p_0$ to a new value $p$ close to $k$ (which from the bulk perspective saves the gradient  energy of the gravitational fields). But deforming the IR lattice period from its original value $p_0$ costs the potential energy among charge carriers. The resulting value of $p$ is then a result of dynamical interplay between these two competing mechanisms. For $p/k =1$ one then has a Mott insulator. 
 
 To find the ratio $p/k$ and construct the full gravity solution requires a tour-de-force numerical analysis of Einstein's equations~\cite{Andrade:2017ghg} (see~\cite{Donos:2012js,Donos:2014uba,Donos:2014oha,Baggioli:2014roa,Gouteraux:2016wxj} for earlier studies of insulators in holography). It was found that for certain ranges of $p_0$, the resulting 
 $p$ is locked into a commensurate state with $k$, i.e. $p/k =1$ and one finds a Mott insulator. Increasing $p_0$ further 
 leads to the resulting $p$ locked into other rational values of $2> p/k > 1$, in which one finds commensurate subregions separated by domain walls, in parallel to the stripes observed in the $La_2CuO_4$ family of cuprate superconductors~\cite{stripes}. 

By changing the temperature one can study the transition from an SLQL phase and the Mott insulator. There exists a critical 
temperature $T_c$, above which one find an SLQL in the presence of a UV lattice and the system is in a metallic phase with 
a  pronounced Drude peak in the optical conductivity. Below $T_c$ the peak of the spectral weight starts shifting to a nonzero frequency and lower the temperature, the larger frequency value of the peak, just as one expects of a Mott insulator.  
Below $T_c$, the resistivity  diverges as a power law $r\sim T^{-1.8}$, rather than  exponentially, as is the case in more conventional Mott systems, which signals a soft gap.

\section{Dynamical  quantum chaos: scrambling \label{sec.scrambling}}

\begin{figure}[t]
\begin{center}
\includegraphics[width=1.0\columnwidth]{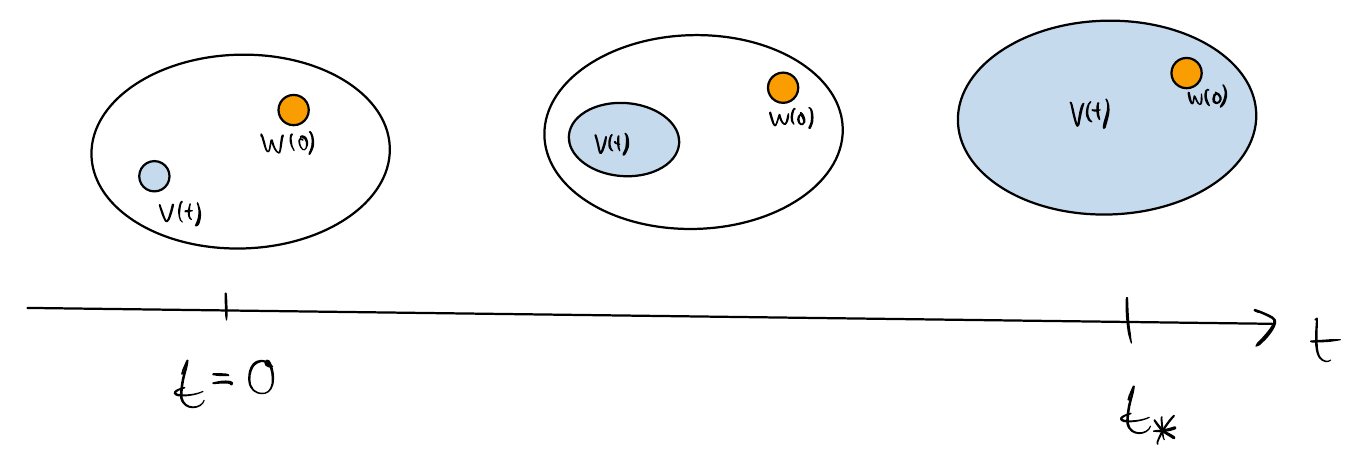}\\
\end{center}
\caption{A cartoon visualization of the scrambling phenomenon. The region inside the circle denotes the space of degrees of freedom including both internal and physical spaces.  A few-body operator $V (t)$ spreads under time evolution 
to encompass a larger and larger region in the space of degrees of freedom until it eventually ``covers'' all degrees of freedom. The time scale for that is called the scrambling time $t_*$. }
 \label{fig:scrambling}
\end{figure}

Scrambling of quantum information has been  increasingly recognized to play a key role 
in  characterizing dynamics of a quantum many-body system. 
Under time evolution, information initially injected into a small subsystem eventually spreads across the entire system. This effect can be described in the Heisenberg picture in terms of growth of the operators (which one uses to inject information), in both physical space and internal space (if there is a large number of local degrees of freedom). 
While for a general many-body system Heisenberg evolutions are very difficult to describe, during the last decade remarkable universalities have been discovered from studies of holographic systems, the SYK models, chaotic spin chains, and random unitary circuits. These studies lead to new approaches to quantum many-body chaos in terms of a quantum Lyapunov exponent, and an emergent 
{butterfly velocity} characterizing the spatial spread of quantum information. They also provide confirmations that black holes are fastest scramblers.

\subsection{Scrambling chaos}

To explore operator growth in both physical and internal spaces, we will have in mind systems with a large internal space,  but the interactions among different degrees of freedom involve only few-body couplings. Examples include holographic systems
for which the internal space consisting of $N \times N$ matrices (with the number of field degrees of freedom $\sN \propto N^2$)
and the SYK models.  

The ``growth'' of an operator can be probed by  \cite{larkin1969quasiclassical,Kitaev-talks:2015}
\be\label{eq.SquaredCommutator}
C(t, \vx) = - \left\langle \left[ V(t,\vx), W(0) \right]^2 \right\rangle_\beta
\ee
where $V$ and $W$ are generic few-body operators which we will take to be Hermitian, and the expectation value is taken with respect to the thermal ensemble at an inverse temperature $\beta$. 
In the large $\sN$ limit, the degrees of freedom involved in $V$ do not overlap with those in $W$. So at $t=0$, the commutator 
between them is very small, with $C(t, \vx)$ of order $1/\sN$. As time increases, $V(t,\vx)$ spreads both in physical and internal spaces, which is reflected in the growth of $C(t, \vx)$.  Eventually the overlap with $W$ becomes $O(1)$ and $C(t, \vx)$ saturates. The time scale of the saturation is often referred to as the scrambling time $t_*$.

$C(t, \vx)$ can also be interpreted as probing the ``sensitivity of a system to initial conditions,'' as follows. 
We can expand the commutator square in~\eqref{eq.SquaredCommutator} and write it as 
\be \label{uhj} 
C_2  = C_1 - C(t, \vx), 
\ee
where (suppressing spatial dependence for notational simplicity) 
\begin{gather} 
C_2 = \vev{\Psi_1 (t)| \Psi_2 (t)} + \vev{\Psi_2 (t)| \Psi_1 (t)}, \\
 C_1 = \vev{\Psi_1 (t)|\Psi_1 (t)} + \vev{\Psi_2 (t)|\Psi_2 (t)} \\
\ket{\Psi_1 (t)} = W(0) V(t) \ket{\Psi_\beta}, \\
 \ket{\Psi_2 (t)} = V(t) W(0)\ket{\Psi_\beta}
\end{gather}
and $\ket{\Psi_\beta}$ denotes the thermal field double state the expectation values with respect to which give the thermal averages. $C_2$ are often referred to as out-of-time-ordered correlators (OTOC) and $C_1$ 
as time-ordered correlators (TOC). $C_2$ describes the overlap between two states $\ket{\Psi_{1,2}}$ which are prepared with slight differences in their initial conditions (whether to first act with the few-body operator $W(0)$). At $t=0$, $C_2$ should be $O(1)$, since few-body operators $V(0)$ and $W(0)$ should commute with each other and thus 
 $\ket{\Psi_{1,2}}$ are almost identical. In a chaotic system, one expects that the slight difference in the initial preparation of $\ket{\Psi_{1,2}}$ will be quickly magnified during time evolution and eventually $C_2$ should decay to zero. In contrast, TOC are simply the norms of $\ket{\Psi_{1,2}}$ and will remain $O(1)$ throughout regardless of whether the system is integrable or chaotic. From~\eqref{uhj}, the growth of $C(t, \vx)$ leads to the decay of $C_2$ and thus the manner of its growth captures the sensitivity to initial conditions of $C_2$. 

In holographic systems, one can use the dual gravity description to compute the OTOCs, which boils down to calculating the  amplitudes for 2-to-2 scatterings of particles dual to $V$ and $W$ in a black hole geometry corresponding to the thermal state~\cite{Shenker:2013pqa}.  One finds, at early times, in {\it all} holographic systems in the classical gravity limit~\cite{Shenker:2013pqa,Roberts:2014isa}  
\be
\label{ehj}
C(t, \vx) \sim \frac{1}{\sN}e^{\lambda \left( t - \frac{|\vx|}{v_B}  \right)} , \quad \lam = {2 \pi \ov \hbar \beta} \ .
\ee
In~\eqref{ehj} $\lam$ can be interpreted as a quantum Lyapunov exponent characterizing how fast the 
operator spreads in internal space and how fast $C_2$ decreases, while $v_B$, often referred to as the butterfly velocity, characterizes {\it ballistic} operator spreading in physical space. Setting~\eqref{ehj} to $O(1)$, one finds that the scrambling time in the internal space is given by $t_* = {\beta \ov 2 \pi} \log \sN$. 

Both $\lam$ and $v_B$ are state-dependent, describing operator growths 
``moderated'' by the state under consideration.  The value of $v_B$ is not universal, depending on details of the dual black hole geometry, which in turn depend on the dynamical exponent,  chemical potentials (if there are conserved charges), or other possible parameters of the boundary system~\cite{Blake:2016wvh,Blake:2016sud}.

The expression for $\lam$ in~\eqref{ehj} is universal for all holographic systems, independent of the details of 
black hole geometries~\cite{Maldacena:2015waa}. It is also not corrected by higher derivative corrections to Einstein gravity with a finite number of derivatives. There is a simple geometric reason for the universality of $\lam$. 
Acting an operator $V$ on the thermal state can be interpreted on the gravity side as adding a particle 
near the boundary of the black hole geometry. The growth of $V(t, \vx)$ with time is reflected on the gravity 
side as the particle moving toward and being accelerated by the black hole. The geometry of near the black hole horizon is such that the energy $E$ of the particle grows with time as $E \propto e^{{2 \pi \ov \hbar  \beta}t}$. The OTOC between $V(t)$ and $W(0)$ is governed by a single graviton exchange between the corresponding particles with an amplitude proportional to $G_N E \sim {1\ov \sN} e^{{2 \pi \ov \hbar  \beta} t}$.

Remarkably, the SYK model and its higher dimensional generalizations have the same behavior for $C(t, \vx)$ as~\eqref{ehj},
but the Lyapunov exponent $\lam$ depends on the temperature in a more complicated way, with the value in~\eqref{ehj} 
obtained in the low temperature limit~\cite{Maldacena:2016hyu,Kitaev:2017awl,Gu:2016oyy}.  
This may be considered as an indication that they are in the same universality class as holographic systems with an Einstein gravity dual.   

The value $\lam$ in~\eqref{ehj} turns out to be special.  It can be shown that the value is in fact the maximal possible value
for an OTOC~\cite{Maldacena:2015waa}, under the assumptions of analyticity and factorization at large times of thermal correlation functions. Thus holographic systems, and SYK models in the low temperature limit are maximally chaotic systems in this sense. This implies that black holes are maximally chaotic and confirms previous expectations that black holes are fast scramblers~\cite{Hayden:2007cs,Sekino:2008he}.

There are systems where different forms of spatial propagation from~\eqref{ehj} were observed, see e.g.\cite{Swingle:2016jdj,Aleiner:2016eni,Patel:2016wdy,Shenker:2014cwa,Gu:2016oyy}, where $C(t, \vx) \sim {1 \ov \sN} e^{\lam t - {|\vx|^2 \ov D_{0} t}}$, described by a diffusive spreading around the exponential growth.

In systems with a small internal space, say a lattice system of qubits, or systems where the interactions in the internal 
space are all to all rather than few-body (e.g. in a random unitary circuit~\cite{Nahum:2017yvy,Nahum:2017yvy,vonKeyserlingk:2017dyr} the scrambling time in the internal space 
is an $O(1)$ number in units of natural time scale of the system. As a result one cannot distinguish a well-defined exponential growth period in $C(t, \vx)$, nor identify unambiguously a Lyapunov exponent.  Nevertheless such systems could still be chaotic based on other measures of quantum chaos such as spectral distributions or the eigenstate thermalization hypothesis (ETH). So the absence of a Lyapunov exponent may not be considered as the absence of quantum chaos. 
Emergence of a butterfly velocity in $C(t, \vx)$ (or equivalently OTOC) indicates ballistic growth of operators in physical space. 

\subsection{Maximal chaos and energy dynamics}

 \begin{figure}
\begin{center}
\resizebox{60mm}{!}{\includegraphics{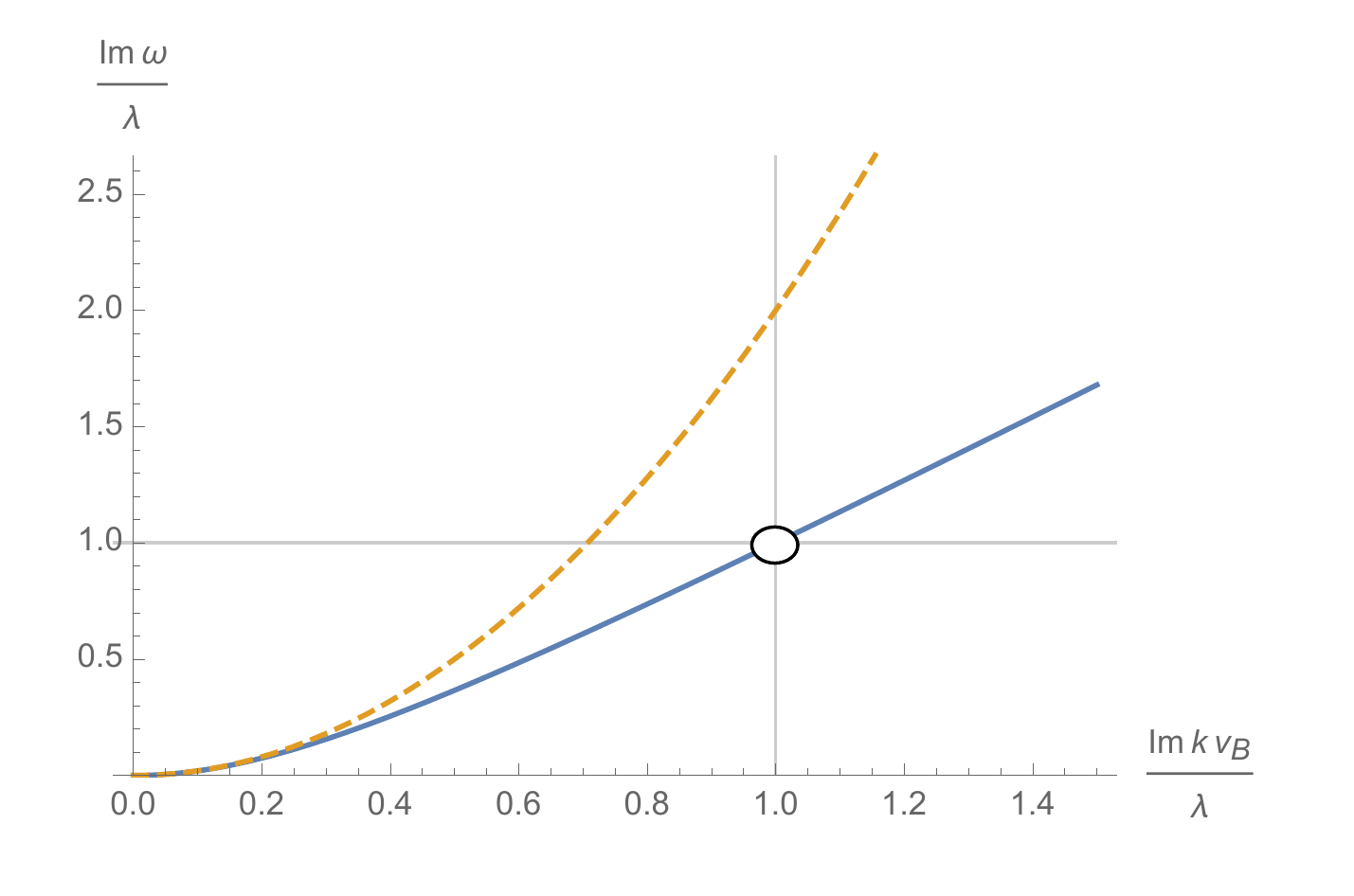}}
\caption{The phenomenon  of  ``pole-skipping'' in the energy-energy correlation function:  following 
the line of poles which starts at small $\om, k$ as the energy diffusion pole along pure imaginary $k$ to a specific value of $k = i \lambda_m/v_B$ for which the pole would be at $\omega= i \lambda_m$, one finds the pole is not there! What happens is that there is also a line of zeros of the energy-energy correlation function precisely passing through that point. 
In the figure the solid line denotes the line of poles, and the open dot indicates that at that particular point the pole is skipped.
The dashed line is the curve $\om = - i D_E k^2$ which coincides with the pole line for 
 small $\om, k$. The line of zeros is not shown. 
 For maximally chaotic systems, pole-skipping gives an alternative way to extract the Lyapunov exponent and butterfly 
 from OTOC.  
}
\label{fig:vanishingpole}
\end{center}
\end{figure}

It is natural to ask whether there is anything special about maximally chaotic systems or 
whether the scrambling chaos has other signatures in addition to those in OTOCs. It turns out 
{\it for maximal chaos}  there is another signature: two-point functions of the energy density exhibit a phenomenon called pole skipping which reflects the non-trivial interplay between hydrodynamic behavior and quantum many-body chaos~\cite{Blake:2017ris,Grozdanov:2017ajz}. 

From studies of the SYK model~\cite{Jensen:2016pah,Maldacena:2016upp} and holographic systems~\cite{Shenker:2013pqa,Shenker:2014cwa} it has been proposed in~\cite{Blake:2017ris} that the growth of a generic operator in internal and physical spaces allows a coarse-grained description in terms of the propagation of an effective field, and furthermore this effective field coincides with the ``hydrodynamic'' field for energy conservation. One can then write down an effective field theory for quantum scrambling chaos by extending the corresponding effective action~\cite{Crossley:2015evo} for energy conservation 
(see also~\cite{Haehl:2018izb} for generalizations).  Pole-skipping is a direct prediction of such an effective description. 

 For illustration, consider a system with energy conservation but not momentum conservation, in which case thermal two-point functions of the energy density exhibit a diffusion pole in momentum space at 
$\om (k) = - i D_E k^2 + O(k^4)$ for $\om, k \to 0$ ($k = |\vec k|$), with $D_E$ the energy diffusion constant.  At finite $\om, k$ the pole 
can be parameterized in terms of a diffusion kernel $\sD (\om, k)$, 
\be \label{yhj}
\om (k) = - i \sD (\om, k)  k^2 , \quad \sD (0,0) = D_E  \ .
\ee
The kernel $\sD(\om, k)$ cannot be determined by the usual arguments of 
hydrodynamics and is in general not known.  The phenomenon of pole-skipping consists of the following two statements:  (1)  $\sD$ satisfies the relation 
\be\label{vb}
 \sD\le(\om_0 ,  k_0 \ri) = {v_B^2\ov \lam_m}  ,  \; \; 
 \om_0 =i \lam_m, \;\;  k_0 = i {\lam_m \ov v_B} 
\ee 
where $\lam_m = {2 \pi \ov \hbar \beta}$ is the maximal Lyapunov exponent. Note that $\om_0 = - i D(\om_0, k_0) k_0^2$
and thus $(\om, k) = (\om_0, k_0)$ lies on the pole line~\eqref{yhj}. (2) At $(\om, k) = (\om_0, k_0)$, the energy-density two-point function also develops a zero, and thus the would-be pole at that point got skipped, see  Fig.~\ref{fig:vanishingpole}.  
For systems with also momentum conservation, instead of a diffusion pole, one finds sound poles and the same pole-skipping phenomenon also happens. Pole-skipping has by now been checked in all currently known maximally chaotic systems including 
the SYK chain~\cite{Gu:2016oyy} in the low temperature limit~\cite{Blake:2017ris},  all finite temperature holographic systems~\cite{Blake:2018leo}, two-dimensional conformal field theories~\cite{Haehl:2018izb}, as well as the Regge regime of conformal field theories~\cite{Haehl:2019eae}.

There are indications that pole-skipping is also present in non-maximal chaotic systems, in which case it captures only the contribution of the stress tensor to scrambling, not the full chaotic scrambling~\cite{MezeiToAppear}.

%It is at the moment an open question whether non-maximal chaotic systems allow an effective field theory description or analogue of the pole-skipping phenomenon. 

 \subsection{Chaos and energy transport}
 
In classical physics chaos plays a key role in transport properties of a statistical system. 
It is natural to ask whether the scrambling-type chaos impacts transport properties of a quantum system. 
For maximally chaotic systems there are both ``empirical'' observations and theoretical arguments 
suggesting that indeed chaos is closely connected to the Planckian dissipation of energy transport. 

Consider a system in which momentum conservation is strongly broken (the energy diffusion constant is infinite with momentum conservation), but energy is conserved. On dimensional grounds, one can write the energy diffusion constant as $D_E \sim {v^2 \ov \tau}$, with $\tau$ and $v$ respectively some characteristic 
time scale and velocity.  For systems without quasi-particles, it appears natural to take $\tau \sim \hbar \beta$~\cite{Hartnoll:2014lpa}, which is also the time scale associated with the maximal Lyapunov exponent $\lam_{m} = {2 \pi \ov \hbar \beta}$. 
For a chaotic system, it has been proposed in~\cite{Blake:2016sud} that $v$ should be taken to be the butterfly velocity 
$v_B$, leading to 
\be \label{yhl}
D_E  = {C \ov 2 \pi} \frac{\hbar v_B^2}{k_B T} = C {v_B^2 \ov \lam_{m}}
\ee
with $C$ some $O(1)$ number. Strong support for~\eqref{yhl} were found for the energy diffusion constant in
all known holographic systems with strongly broken momentum conservation~\cite{Blake:2016wvh,Blake:2016sud}. 
In many examples, both $D_E$ and $v_B^2$ often each have complicated dependence on various 
parameters of a system, but miraculously their ratio becomes very simple, given by some $O(1)$ constant 
times $\hbar \beta$ as in~\eqref{yhl}. 

The relation~\eqref{yhl} can be argued as a consequence of the presence of an effective field playing the dual 
 role of energy conservation and operator growth. More explicitly, with diffusion kernel $\sD (\om, k)$ being a smooth function of $\om$ and $k$, one expects $C = {\sD (\om=0, k=0) \ov \sD\le(\om =i \lam_m,  k = i {\lam_m \ov v_B} \ri)}$ to be an $O(1)$ number. Equation~\eqref{yhl} then immediately results from the second equation of~\eqref{yhj} and~\eqref{vb}. 

For charge diffusion, however, the picture is more complicated \cite{Lucas:2016yfl,Davison:2016ngz,Zhang:2020hzz} which does not appear to have a direct relation with chaotic dynamics in the sense of scrambling.
\begin{figure}[t]
\begin{center}
\includegraphics[width=.7\columnwidth]{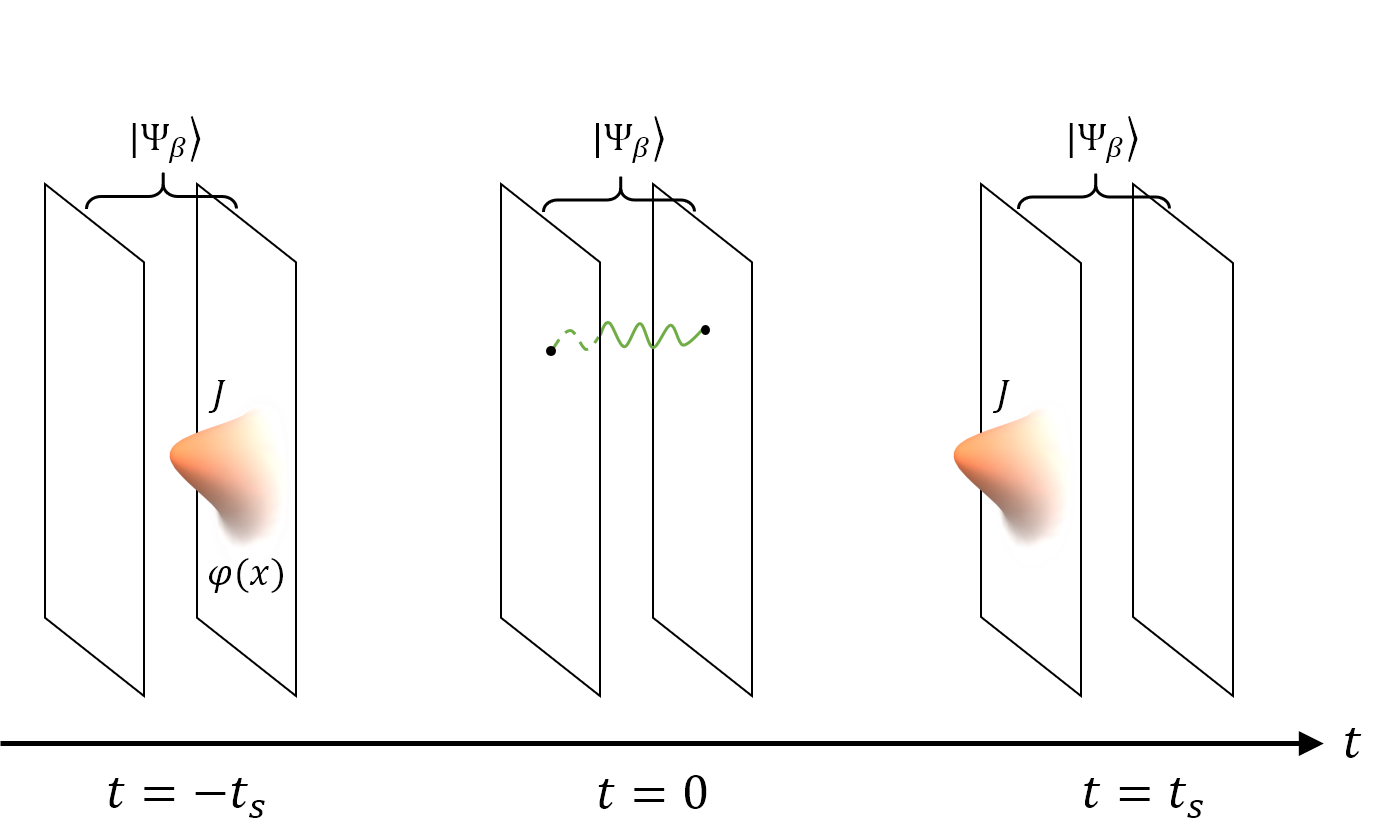} \includegraphics[width=.6\columnwidth]{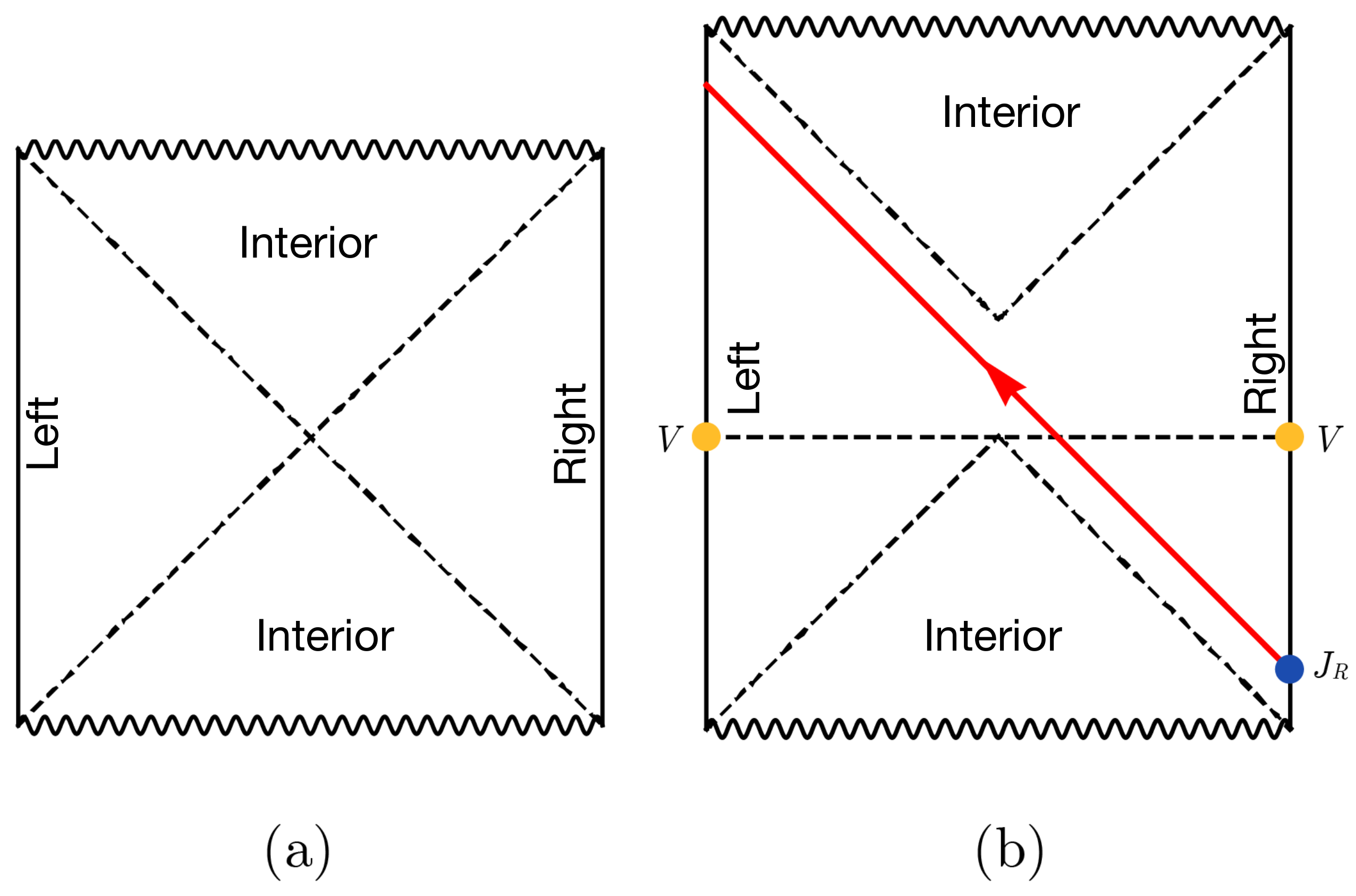}  \\
\end{center}
\caption{{\bf Up}: we prepare two identical copies of the same system in a thermal field double state.  
At  time $-t_s$ an input signal is introduced into the right copy. A coupling between two systems is activated at time $t=0$.
This causes the scrambled signal to reconstitute spontaneously in the left system at $t=t_s$. {\bf down}: 
(a) the causal structure of an eternal black hole. The left and right boundaries are connected by the Einstein-Rosen bridge, which is non-traversable. Lights propagate along 45 degree lines. The top and bottom wavy lines are the black hole singularities. 
Any signals sent from left or right will fall into the top singularity and cannot reach the other side. (b) The backreaction of the coupling between left and right systems at $t=0$ deforms the geometry and the causal structure. As a result,  if sent early enough, 
a signal from the right side can reach to the left side where it reappears. For an observer on the right, the signal simply dissipated.}
 \label{fig:GJWwormhole}
\end{figure}

\subsection{Regenesis and a traversable wormhole} 

As another illustration of the physical implications of scrambling chaos we discuss a surprising phenomenon called ``regenesis'' 
which arises from a nontrivial interplay between scrambling and entanglement.  The phenomenon also highlights the fact that when  
a disturbance dissipates as in Fig.~\ref{fig:NoQuasiParticle} it is not really lost, but just scrambled. Under certain protocol, it can be brought back. This phenomenon was first discovered 
on the gravity side as a way to generate traversable wormholes using quantum effects~\cite{Gao:2016bin,Maldacena:2017axo}, but can be shown to be universal in general quantum chaotic systems~\cite{Gao:2018yzk}.

Let us consider two copies of a chaotic system, with the total Hamiltonian $H_0 = H^L + H^R$, where $H^{L,R}$ are respectively the Hamiltonian for the two copies, one referred to as the `left' and the other as the `right' system.  One prepares the total system in a special entangled state: the thermofield double state at inverse temperature $\beta$. Suppose at time $-t = t_s > t_*$ ($t_*$ is the scrambling time) one slightly disturbs the right system, the signal of which quickly dissipates after a short time, typically of order $\hbar \beta$. 
Long after the signal has disappeared, one turns on a coupling between the two copies for a short while at $t=0$,
which we can approximate as a delta function in time. In other words the full Hamiltonian has the form $H = H_0 - g  V \delta(t=0)$. The coupling $V$ takes the schematic form ${\cal V}^L {\cal V}^R$, where $\sV$ is some general few-body operator. One then finds that at $t=t_s$, the signal reappears in the $L$ system. The reappeared signal is not identical to the original signal, but related by a transformation. See Fig.~\ref{fig:GJWwormhole}. 
While at first sight this ``regenesis'' is rather counter-intuitive,  it can be shown to be a simple consequence of two elements:  
 (i) scrambling which in a chaotic system makes  (OTOCs) vanish at large times; (ii) the entanglement structure of the thermofield double state.  Heuristically, the quantum information of the original signal is scrambled among the right system, and then gets transmitted to the left system through the coupling at $t=0$, finally is reconstructed at time $t_s$ due to the entanglement structure between the left and right systems. Chaos endures that the phenomenon is robust for generic
choices of the operator $\sV$ regardless of the nature of the original signal. 
 
 Compared to~\eqref{ehj} which deals with early times, regenesis is concerned with the time scales of order the scrambling time of a system. 
 
For maximally chaotic systems with a gravity dual, the phenomenon has a beautiful geometric interpretation. The thermal field double state corresponds to an extended Schwarzschild geometry, which has a non-traversable wormhole, the so-called Einstein-Rosen bridge. Turning on the coupling potential $V$ at $t=0$ injects a pulse of negative energy which makes the Einstein-Rosen bridge traversable: the signal sent much earlier in the right system can now simply propagate through the wormhole to show up in the left system. See Fig.~\ref{fig:GJWwormhole}. 

Connecting quantum chaos on the field theory side and interesting gravitational phenomena in this way, gives a new and exciting perspective on potential laboratory realizations of quantum many-body systems with holographic duals. Evidently as the relevant technology is at an early stage, one must initially target the  simplest known realizations, namely SYK-like systems. For early implementations in cold-atomic gases as well as approaches using digital quantum simulation, see \cite{Danshita:2016xbo,Garcia-Alvarez:2016wem,Franz:2018cqi}. A more recent proposal for harnessing such platforms in order to study the regenesis phenomenon in the laboratory appears in \cite{Brown:2019hmk}. All these works underline the point that holographic dualities offer what is arguably the most promising avenue to the study of analog quantum gravity systems in the laboratory.

\subsection{Other notions of quantum chaos}

The scrambling chaos discussed above takes place at time scales up to the scrambling time, which for 
a large $\sN$ system, scales as $\log \sN$. 
There are two other notions of quantum chaos, based on 
energy level distributions (spectral chaos) and properties of energy eigenstates. 
Results on these subjects from holography are somewhat scarce, so we will be brief.

\begin{figure}[t]
\begin{center}
\includegraphics[width=.8\columnwidth]{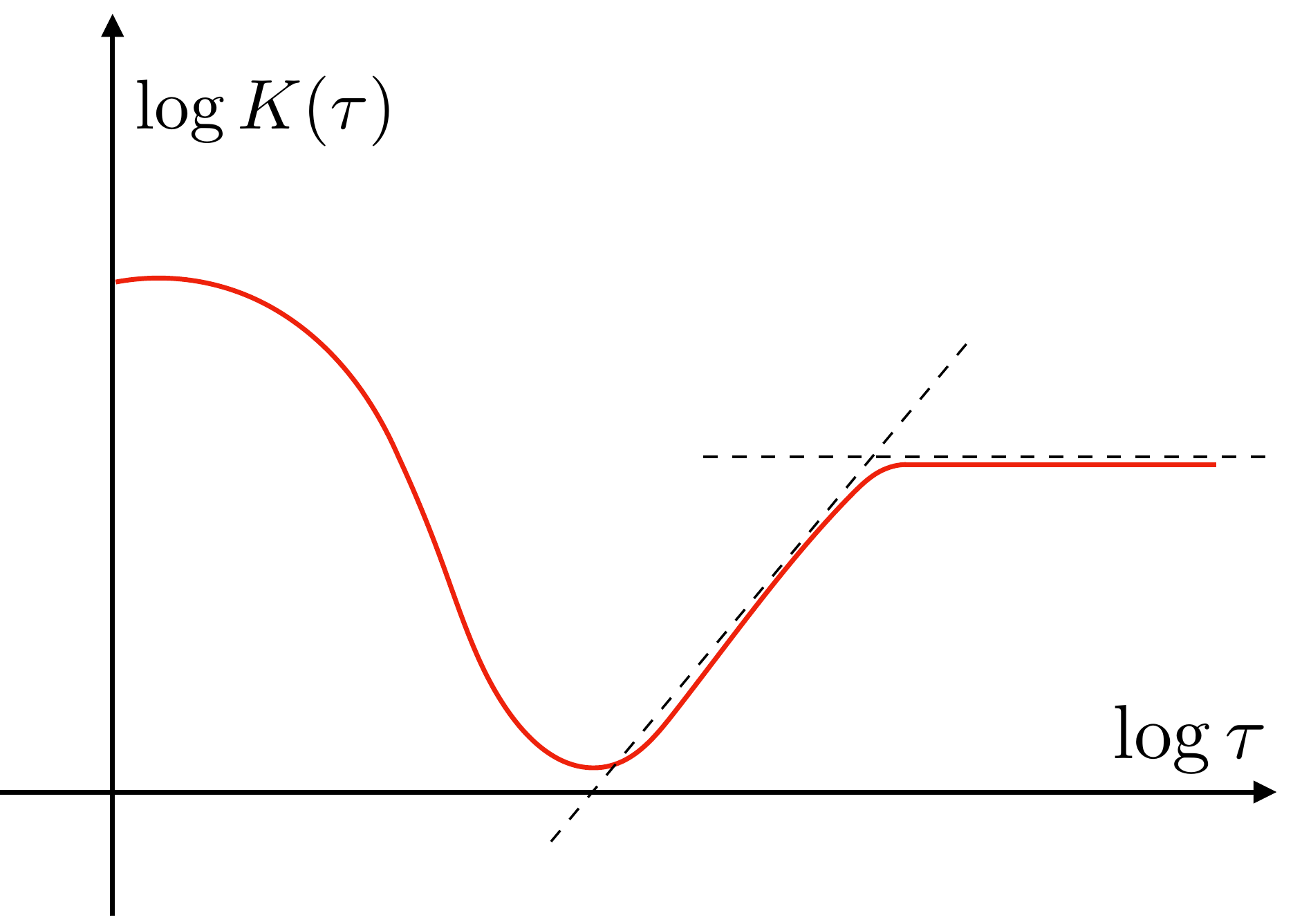}  \\
\end{center}
\caption{Universal behavior of $K (\tau)$ for a chaotic system~\cite{haake1991quantum}. $\tau$ is measured in units of $\Delta^{-1}$, where $\De$ is average level spacing. The universal RMT behavior corresponds to the linear ramp followed by a plateau, both indicated in dotted lines in the figure.}
 \label{fig:ktau}
\end{figure}

One criterion for distinguishing a quantum chaotic system from integrable ones is that energy levels should show level repulsion, or more precisely, the distribution of spacings of energy levels should show random-matrix like behavior~\cite{bohigas1984characterization,mehta2004random}.
For a system with $\sN$ degrees of freedom, on general ground the average level spacing $\De$ scales as 
$\De \sim e^{-\sN}$. So the spectral chaotic behavior is relevant for physics at very long time scales 
of order $\De^{-1} \sim e^{\sN}$.  An important ``observable'' to  probe the level distribution is the spectral form factor $K(\tau)$, which is defined as the Fourier transform with respect to $\om$ of the level-density two point function $\langle \rho(E) \rho(E+\om) \rangle$, where $\rho (E)$ is the density of states and $\vev{\cdots}$ denotes averaging over a band of energy levels. See e.g.~\cite{haake1991quantum} for more details.
For a chaotic system, $K(\tau)$ has the universal behavior shown in {Fig.~\ref{fig:ktau}}, with a ramp and a plateau.

As discussed in previous subsections, holographic systems exhibit maximal scrambling chaos, it is then a natural to pose the question whether holographic systems also exhibit spectral chaos, in particular, whether one can derive the universal behavior of $K(\tau)$ using gravity. This latter question is much deeperxs than just confirming spectral chaos, but touches essential non-perturbative 
aspects of quantum gravity. 
 Spacetime in general relativity and in perturbative string theory is continuous. But from holographic duality, quantum gravity is dual to quantum systems with discrete energy levels. How such discreteness shows up on the gravity side and its implications 
 for spacetime structure are outstanding open questions whose answers should provide key insights into the fundamental nature of quantum gravity. $K(\tau)$ is the simplest observable to probe such discreteness. 

Recently there has been exciting progress in obtaining the ramp behavior in $K(\tau)$
in a two-dimensional gravity theory, called Jackiw-Teitelboim (JT) gravity theory~\cite{Jackiw:1984je,Teitelboim:1983ux}. The JT gravity theory is very similar to the standard Einstein action, up to the addition of the scalar degree of freedom $\phi$, which is necessary to obtain an interesting two-dimensional theory. It is shown 
in~\cite{Stanford:2019vob,Stanford:2019vob,Saad:2019lba} that the expansion of certain path integral in the JT gravity theory in terms of spacetime topologies to be summed over precisely reproduces the $1/D$ expansion of level-density two point function $\langle \rho(E) \rho(E+\om) \rangle$ in a random matrix theory, where $\vev{\cdots}$ is now averaged over random Hamiltonians drawn from certain ensembles and $D$ is the rank of ransom matrices. A Fourier transform of the expansion to obtain $K(\tau)$ leads to the ramp behavior, while the plateau remains illusive from the gravity perspective, although important hints have been gained how it may arise in principle \cite{Stanford:2019vob}. 

A surprising feature of this study is that while the JT gravity is a fixed gravity theory, 
it turns out to be dual to an average over an ensemble of random Hamiltonians. This is distinct from the usual situation
in holographic duality, with the gravity theory dual to a boundary system with a fixed Hamiltonian.  
Generalization of the discussion of~\cite{Stanford:2019vob,Saad:2019lba} to higher dimensional gravity theories appear
daunting at the moment, nor is it clear how random matrix behavior for a boundary system with a fixed Hamiltonian can 
arise from gravity.

Another manifestation of quantum many-body chaos is the eigenstate thermalization hypothesis (ETH) which says that expectation values of general few-body operators in a sufficiently excited energy eigenstate should behave like those in a thermal state~\cite{PhysRevA.43.2046,Srednicki}. ETH is consistent with the general expectation of holographic duality: a generic highly excited state is believed on the gravity side to be described by a black hole, which exhibits thermal behavior.
However, the explicit construction of gravity geometry dual to a single energy eigenstate is still out of reach, and so is a direct verification 
that holographic systems satisfy ETH.  Nevertheless some recent progress has been made in two-dimensional and three-dimensional gravity theories, see e.g.~\cite{Fitzpatrick:2014vua,Fitzpatrick:2015foa,Lashkari:2017hwq,Basu:2017kzo,Brehm:2018ipf,Das:2017cnv,Lam:2018pvp,Nayak:2019evx,Saad:2019pqd,Kraus:2016nwo}.

\section{Quantum information and spacetime structure} \label{sec:QI}

. 

The last decade has seen many exciting developments revealing deep connections between spacetime geometry and quantum information of the boundary systems, pioneered by the Ryu-Takayanagi formula~\eqref{eq.RTformula}. 
The R-T formula and its generalizations provide powerful ways to compute entanglement entropies in strongly coupled systems which are otherwise impossible to obtain. 
This led to many new insights into equilibrium and non-equilibrium dynamics of a quantum many-body system, such as 
the structure of entanglement of a quantum field theory (see e.g.~\cite{Ryu:2006bv}), ``evolution'' of entanglement under renormalization group flows (see e.g.~\cite{Myers:2010xs,Myers:2010tj,Liu:2012eea,Liu:2013una}), entanglement growth during thermalization (see e.g.~\cite{Hubeny:2007xt,AbajoArrastia:2010yt,Albash:2010mv,Balasubramanian:2010ce,Galante:2012pv,Caceres:2012em,Arefeva:2013wma,Hartman:2013qma,Liu:2013iza,Anous:2016kss})
and more recently the black hole information loss paradox~\cite{Penington:2019npb,Almheiri:2019psf,Almheiri:2019hni}.
This is a big subject, for which extensive reviews of various topics already existe~\cite{Rangamani:2016dms,Liu:2018crr}.
The quantum information perspective also led to deeper insights
into the structure of holographic duality itself, including the formulation of subregion duality and 
the realization of states in the boundary system dual to bulk geometries as quantum error correction codes.   
Here we will highlight some recent developments in this direction including progress in understanding the unitarity of black hole evaporation (see also~\cite{Harlow:2014yka,Harlow:2018fse,Maldacena:2020ady} for earlier reviews).

\subsection{Subregion duality and quantum error correction}

The R-T formula~\eqref{eq.RTformula} expresses the entanglement entropy $S (\rho_A) \equiv - \Tr_A \rho_A \log \rho_A$ of a boundary subregion $A$ in terms of a classical geometric quantity (the area of a minimal surface) on the gravity side, and is valid to  leading order in the $1/\sN$ expansion (or $\hbar G_N$ expansion). Quantum effects on the gravity side appear as 
 higher order corrections. It turns out there is a simple way to take account of the next order corrections~\cite{Faulkner:2013ana}, 
\be \label{hrn}
S (\rho_A)= \frac{{\rm Area}(\ga_A)}{4\hbar G_N} +  S (\rho_\fa) + O(\hbar G_N) \,,
\ee
where $\rho_\fa$ and $S (\rho_\fa) \equiv - \Tr_\fa \rho_\fa \log \rho_\fa$ are respectively the reduced density matrix and the bulk entanglement entropy in the region $\fa$ between the minimal surface $\ga_A$ and $A$, see Fig.~\ref{fig:RTEntanglementWedge}. We have used $\rho$ to denote both the states of the bulk gravity and boundary theories as they are identified. On the gravity side it includes not only the classical geometry but also quantum fluctuations around the geometry. $S (\rho_\fa)$ can be expanded in $\hbar G_N$, with the leading contribution, of order $(\hbar G_N)^0$,  given by the entanglement entropy of the graviton and matter fields in~\eqref{nmb} by treating them as quantum fields in a fixed spacetime geometry. 

\begin{figure}[!h]
\begin{center}
\includegraphics[width=3.5cm]{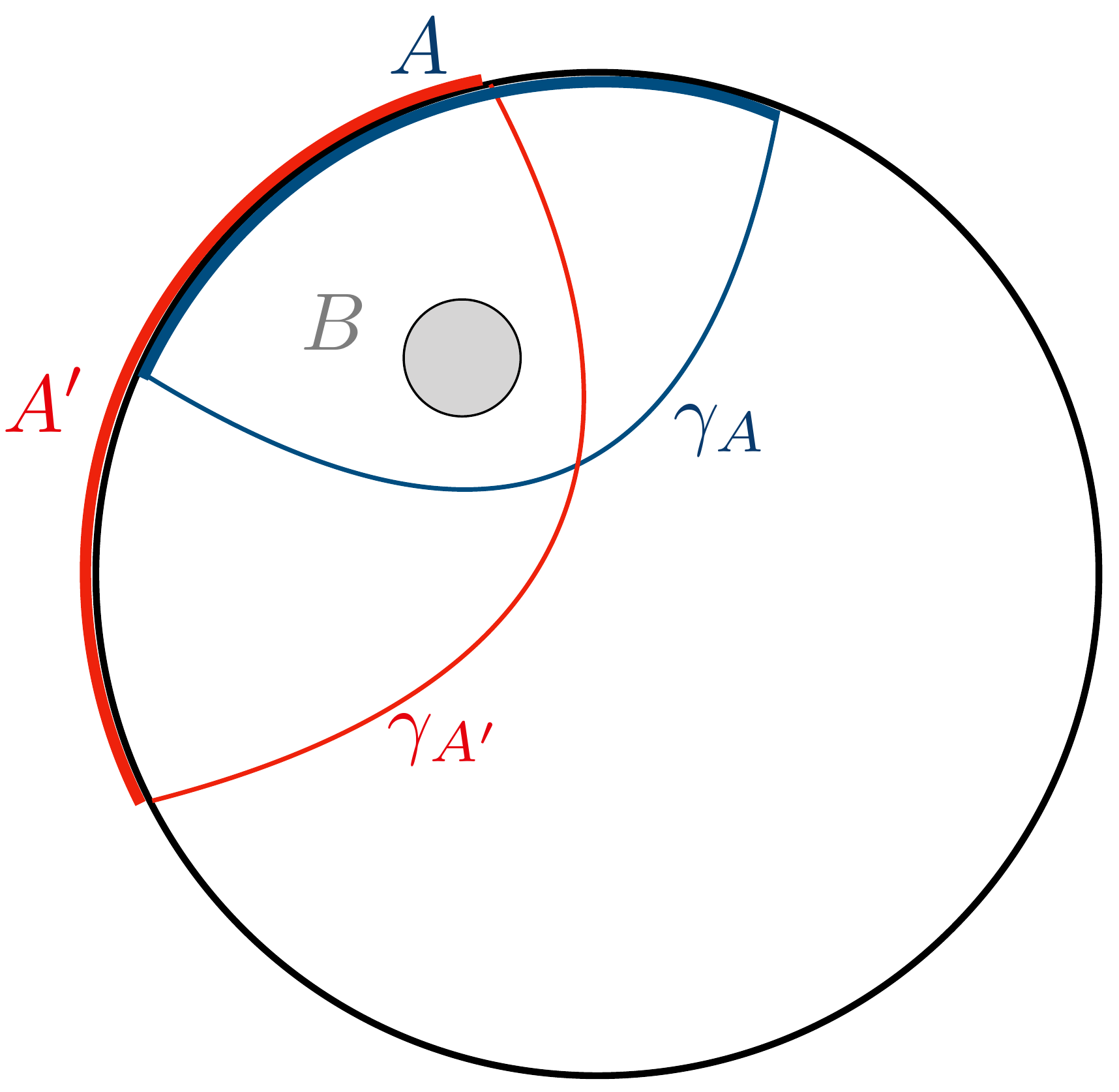}
\caption{Quantum information in a local subregion $B$ is encoded in the boundary system in a highly redundant way as 
it can be contained in the entanglement wedges of many possible boundary subregions. Here the figure should be understood as representing a single time slice in AdS with the circle denoting the boundary while the interior of the circle bulk of AdS.
A bulk subregion $B$ is shown to be included in the entanglement wedges of boundary subregions $A$ and $A'$, 
and in principle infinitely many others. 
}
\label{fig:redu}
\end{center}
\end{figure}

Equation~\eqref{hrn} is physically intuitive; it says that the leading quantum corrections to $S (\rho_A)$ is given by 
the bulk entanglement entropy in region $\fa$, often referred to as the entanglement wedge of $A$. 
Even though the second term in~\eqref{hrn} is much smaller than the classical contribution in the semi-classical limit 
$\hbar G_N \to 0$, it has important conceptual implications. It gives special physical significance to the entanglement wedge $\fa$ as the bulk subregion corresponding to the boundary subregion $A$. 
Consider, for example, adding a qubit to region $\fa$ on the gravity side. 
The bulk entanglement entropy $S_\fa$ depends on whether and how the qubit is entangled with other degrees of freedom, and from~\eqref{hrn}  so does  the boundary entropy $S_A$.
But adding a qubit to the complement $\bar \fa$ will not lead to any change in $S_A$. This intuitive picture can be made mathematically precise by using the relative entropy.  Let us consider another state $\sig$ which is close 
to $\rho$ in the sense that $\sig$  may be considered as adding a finite number of graviton or matter excitations to $\rho$.
Then one can show from~\eqref{hrn} that~\cite{Jafferis:2015del} 
\be \label{inj0}
S(\rho_A||\sig_A) =  S (\rho_\fa ||\sig_\fa)  + O(\hbar G_N)
\ee
where $S(\rho_A||\sig_A) = {\rm Tr}  (\rho_A \log \rho_A - \rho_A \log \sig_A)$ is the relative entropy between $\rho_A$ and $\sig_A$. The relative entropy can be considered as a measure of the distinguishability of
two quantum states. In particular, when $\rho_A = \sig_A$,  $S(\rho_A||\sig_A) =0$, then equation~\eqref{inj0} implies $\rho_\fa = \sig_\fa$ (up to corrections in trace distance of $O(\hbar G_N)$), and vice versa. This implies that there should be  a one-to-one map between the reduced density matrices $\rho_\fa$  and $\rho_A$.  In other words, knowledge of one could be used to determine the other, which is often referred to as subregion duality~\cite{Czech:2012bh,Wall:2012uf}. Indeed it can be shown~\cite{Dong:2016eik} from~\eqref{inj0} that the action of any bulk operator localized in $\fa$ on a state $\rho$ can be reproduced by the action of a boundary operator localized in region 
$A$~(see~\cite{Hamilton:2006az,Morrison:2014jha} for earlier work). 

The subregion duality has a beautiful interpretation from 
quantum information perspective in terms of quantum error correction, i.e. the quantum state dual to a semi-classical bulk geometry can be understood as a quantum error correcting code~\cite{Almheiri:2014lwa}. 
Heuristically, that one can recover all the quantum information in the bulk subregion $\fa$ with knowledge of only that in 
the boundary subregion $A$ implies that the information in $\fa$ is robust even if we ``make errors'' in or even completely ``erase''  the complement $\bar A$ of $A$. Furthermore, any quantum information in a local bulk region is stored in the state in a highly redundant way, see Fig.~\ref{fig:redu}.  Mathematically, given a state $\rho$, one can define a quantum (error) channel $\sE$ as $\sE (\rho) = {\rm Tr}_{\bar A} \rho$  (tracing over $\bar A$  causes errors). The one-to-one mapping between $\rho_A$ and $\rho_\fa$ implies the existence of a recovery channel $\sR$ which gives $\sR (\sE(\rho)) = {\rm Tr}_{\bar \fa} \rho$~\cite{Cotler:2017erl}. The error correction perspective is powerful since  the generalized R-T formula~\eqref{hrn}, the relation between bulk and boundary relative entropies~\eqref{inj0}, and subregion duality can all be thought of as different facets of the quantum state of the system being a quantum error correcting code~\cite{Harlow:2016vwg}. 
 
We should note the error correction is approximate at this level as there are higher order corrections in~\eqref{inj0}, which 
also has important implications~\cite{Cotler:2017erl,Hayden:2018khn}.

\subsection{Quantum extremal surface} 

In usual situations the quantum bulk entanglement entropy term in~\eqref{hrn} is much smaller than the classical term. 
But there are situations, such as the evaporation of a black hole (as we will see below), where
the quantum contribution can become comparable to the classical contribution even in the semi-classical $\hbar G_N \to 0$ limit.
In such cases equation~\eqref{hrn}  breaks down and one should reorganize the $\hbar G_N$ expansion. 
Sometimes even with the second term much smaller than the classical term, it can happen that the spacetime gradient of
$S (\rho_\fa)$ becomes comparable with that of the classical term, in which case it is also questionable whether one should  extremize only the classical term. It turns out there is a simple remedy~\cite{Engelhardt:2014gca}. A hint is that in~\eqref{hrn} the two terms are treated unevenly: one first extremizes the area of  
codimension-two surfaces to find $\ga_A$ and then use $\ga_A$ to define the entanglement wedge $\fa$ to find $S_\fa$.  
But there are physical reasons that the two terms should be treated on equal ground. 
 
To see this, let us first review the concept of a generalized entropy. Consider a spacelike, codimension-$2$ surface $X_A$ which is homologous to $A$, and $\p X_A = \p A$. We will denote the region between $X_A$ and $A$ on a Cauchy slice again to be $\fa$ (i.e. $\p \fa = A \cup X_A$), and the entanglement entropy of the bulk system in region $\fa$ as $S_\fa$. One then defines a generalized entropy associated with $X_A$ as 
\be \label{hrn0}
S_{X_A} = \frac{{\rm Area}(X_A)}{4\hbar G_N} + S_\fa \ .
\ee
We note that from unitarity $S_\fa$ is independent of the Cauchy slice one uses to define $\fa$. 
The generalized entropy~\eqref{hrn0} was the key to generalizing the second law of thermodynamics 
to systems with a black hole where $X_A$ is taken to be the event horizon~\cite{Bekenstein:1973ur,Hawking:1974sw}.
Furthermore, there are reasons to believe that the combination~\eqref{hrn0} is UV finite (see e.g.~\cite{Bousso:2015mna}, which contains a review of the situation).  
Thus the two terms in~\eqref{hrn0} should be viewed as different parts of the same  ``entropy property'' associated with $X_A$.  With this understanding in mind, the generalization of~\eqref{hrn} is then immediate; to obtain $S(\rho_A)$ we simply extremize the full generalized entropy with respect to $X_A$~\cite{Engelhardt:2014gca} 
\be  \label{hrn1}
S_A = S_{\ga_A}  = {\rm extremize}_{X_A} S_{X_A} 
\ee
where the extremum surface $\ga_A$ is now called the quantum extremal surface. 
If there are multiple quantum extremal surfaces, then one picks the one with the smallest generalized entropy. 

When $S_\fa$ is of order $(\hbar G_N)^0$, at leading order $\ga_A$ is given by the classical extremal surface, and~\eqref{hrn1} recovers~\eqref{hrn} at order $(\hbar G_N)^0$. It can in fact be used to generalize~\eqref{hrn} to all orders in $(\hbar G_N)$ expansion when one also generalizes the classical area term to that appropriate for higher derivative gravity corrections~\cite{Engelhardt:2014gca,Dong:2013qoa,Dong:2017xht}.

\begin{figure}[!h]
\begin{center}
\includegraphics[width=5cm]{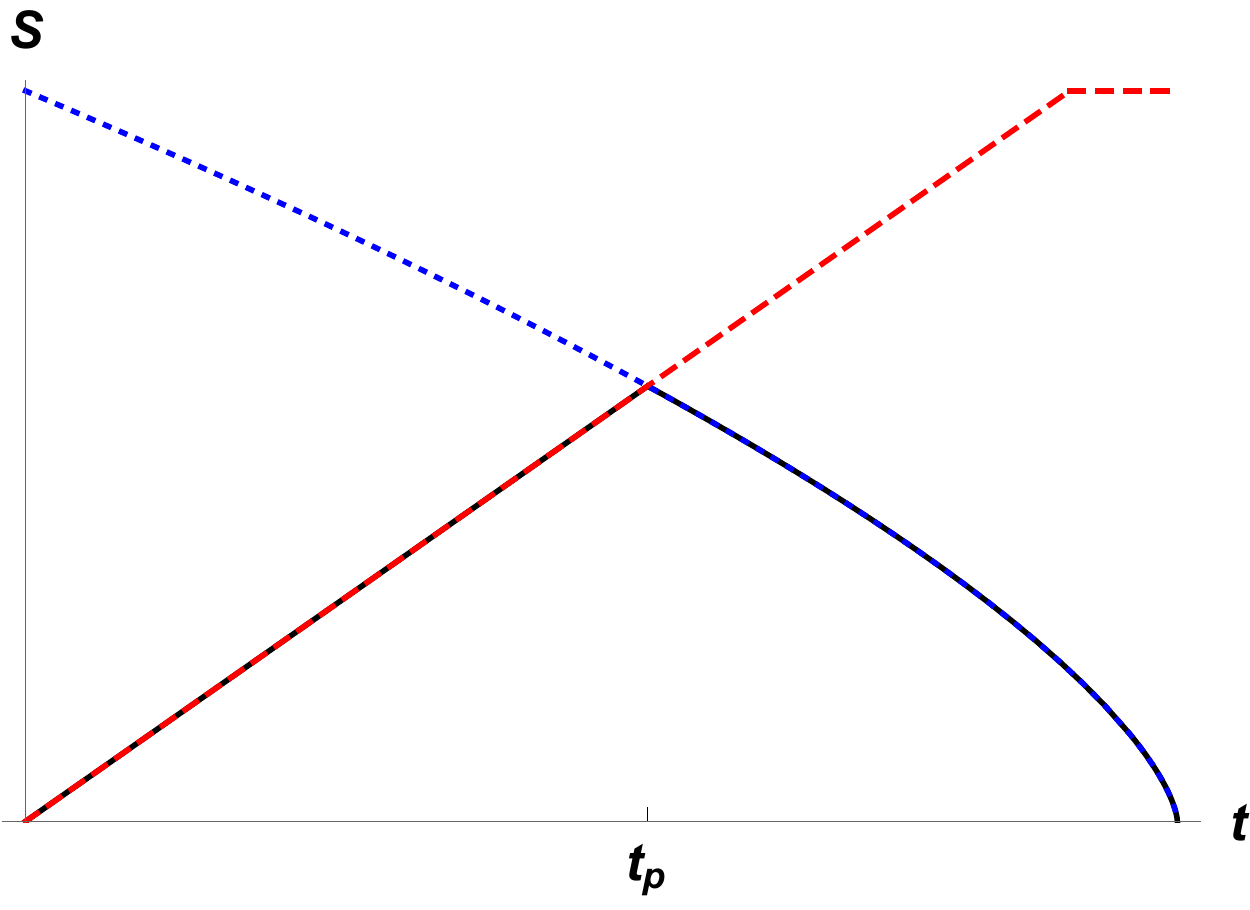}
\caption{The Page curves of the black hole and the radiation. At $t=0$, the black hole consists of the whole system and is in a pure state. The dotted line is the semi-classical entropy of the black hole from its horizon area. The dashed line is the entropy of the semi-classical radiation from Hawking's calculation. The solid curve is the entanglement entropy for the black hole and the radiation in a full quantum description. It should be seen as two curves, one for the black hole and one for the radiation, which coincide as required by unitarity.  The Page time $t_p$ refers to time scale where the solid curve  turns around from increasing to decreasing with time. The details of the curve are not important other than the curve should first monotonically increase and then monotonically decrease. 
}
\label{fig:page}
\end{center}
\end{figure}

\subsection{An application to black hole information loss paradox}

As an example of the application of the quantum extremal surface and subregion duality, we discuss how they can be used to shed light on the black hole information loss paradox. 

Hawking's observation that a black hole emits thermal radiation appears to violate unitarity of quantum mechanics~\cite{Hawking:1974sw,Hawking:1976ra}: when a black hole in a pure state evaporates completely, one is left with only the emitted thermal radiation, which naively is in a mixed state. 
{It has long been recognized that Hawking's observation is not really in contradiction with unitarity as
possible global quantum correlations among different parts of the Hawking radiation which purify  the final state
may not be visible in Hawking's semi-classical calculations. 
Later Page~\cite{Page:1993wv} pointed out a highly nontrivial test for the unitarity of evaporation process:}
the entanglement entropy of the radiation and the black hole must be equal, and their time evolution must follow what is now widely referred to as the Page curve, initially rising monotonically until the so-called Page time and then monotonically decrease  back to zero, see Fig.~\ref{fig:page}.  

The argument for the Page curve is very simple. Consider a quantum system $L = B \cup R$ whose Hilbert space decomposes as $\sH_L = \sH_{B} \otimes \sH_{R}$, with the dimensions of $\sH_{B} $ and $\sH_{R} $ respectively equal to $d_{B}$ and $d_{R}$. One expects for a system in a typical pure state~\cite{lubkin,Lloyd:1988cn,Page:1993df} the entanglement entropy for $S_B, S_R$ for the $B$ and $R$ subsystems is given by (assuming both $d_B, d_R$ are large and not equal) 
\be \label{avEn}
S_B = S_R = {\rm min} (\sS_B, \sS_R) + \cdots 
\ee
where $\sS_{B,R} = \log d_{B,R}$ are the ``coarse-grained entropies'' of the $B$ and $R$ subsystems. 
For an evaporating black hole, one takes $B$ and $R$ to be the black hole and radiation subsystems respectively. {The Hilbert space of both $B$ and $R$} changes with time. At $t=0$, $B (t=0) = L$ while $R(t=0)$ is empty, and as time goes on the degrees of freedom in $B(t)$ slowly go over to $R(t)$, until the black hole has completely evaporated. The Page curve then follows from~\eqref{avEn}, see 
{Fig.~\ref{fig:page}}. The turnaround time scale is often referred to as the Page time, which happens when $d_{B(t)} \sim d_{R(t)}$.

In the usual semi-classical treatment \`a la Hawking, the Page curve is not observed. The entropy of the black hole is given by its horizon area which decreases over time monotonically, while the entropy of the radiation increases monotonically
until the completion of the evaporation, as indicated in Fig.~\ref{fig:page}. There is no real contraction with unitarity as these entropies are coarse-grained thermodynamic entropies rather than the fine-grained entanglement entropies. 
It had often been believed that a derivation of the Page curve may require finer control of a gravity system than the semi-classical approximation. It thus came as a pleasant surprise when recently the Page curve was derived 
from a semi-classical treatment based on~\eqref{hrn1}~\cite{Penington:2019npb,Almheiri:2019psf}, 
and highlights the power of~\eqref{hrn1} to capture fine-grained quantum information of gravity.
Here we will outline the basic ideas and physical intuitions of their derivations without going into technical details of the specific setups and calculations.  

Consider a black hole formed from gravitational collapse of some matter in a pure state in AdS. With the standard reflecting condition at the boundary of AdS, a sufficiently large black hole will come to be in thermal equilibrium with its own Hawking radiation and will be there forever. Now let us imagine modifying the boundary condition to a free propagating one in which case the Hawking radiation will leave AdS and the black hole will eventually evaporate. The full system can be described as 
$L = B(t) \cup R(t)$ with $\sH = \sH_{B(t)} \otimes \sH_{R(t)}$. Here the subsystem $B(t)$ describes quantum gravity in AdS (including the black hole)  and is described by some boundary many-body system. The subsystem $R(t)$ includes all the Hawking radiation which has left the AdS space. The division of $L$ into $B(t)$ and $R(t)$ depends on time as degrees of freedom escape from the black hole in AdS to $R$ as Hawking radiation.

\begin{figure}[t]
\begin{center}
\includegraphics[width=.8\columnwidth]{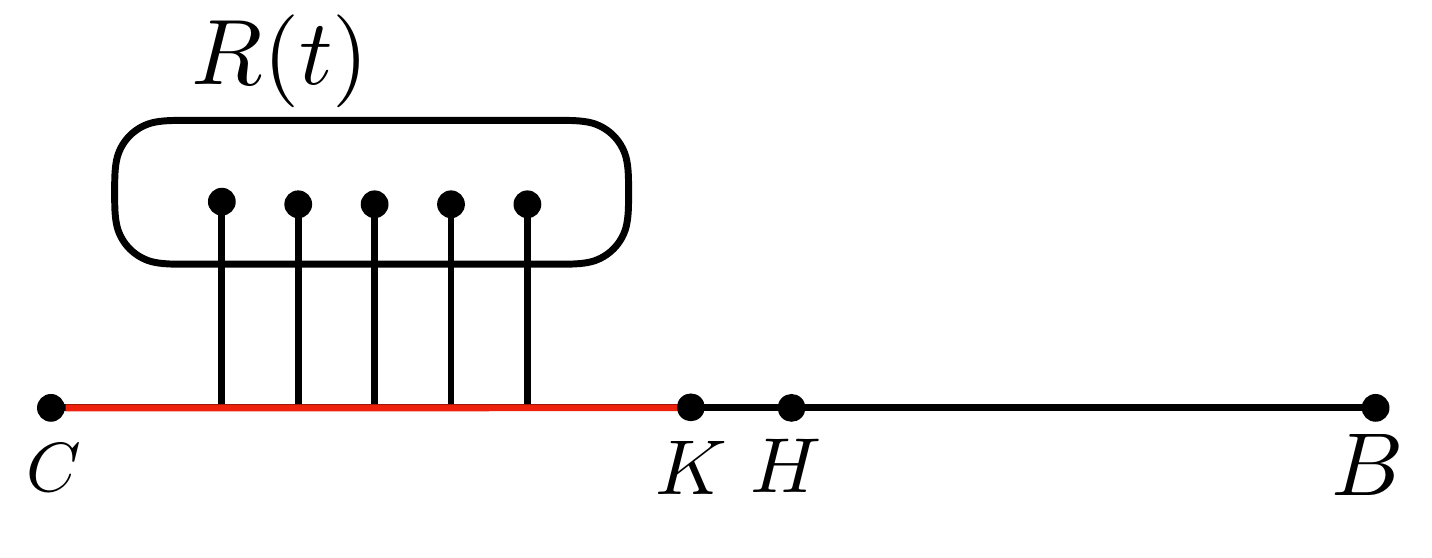}\\
\end{center}
\caption{Cartoon of a black hole in AdS and the Hawking radiation. The line in the figure represents  a single time slice in AdS with each point on the line representing a codimension-2 surface which is homologous to the boundary spatial manifold. $B$ denotes the boundary, $H$ the horizon, and $C$ the center of AdS. The region between $C$ and $H$ is the interior of the horizon which is entangled with radiation subsystem $R(t)$ due to the Hawking process, with entanglement shown using vertical lines between $R(t)$ and the black hole interior. The line between $H$ and $B$ is region in AdS outside the black hole. 
Before the Page time, the quantum extremal surface is at $C$ with zero area. 
$K$ is the location of quantum extremal surface after the Page time, which is slightly inside the horizon (distance of $KH$ is proportional to $G_N$).
The region $CK$ (which is colored red), denoted as $\sI$ in the main text, belongs to the entanglement wedge of the radiation $R(t)$.
 }
 \label{fig:BH}
\end{figure}

One important element to keep in mind is that the Hawking process can be considered as a pair creation process near the horizon. The created pair is entangled,  with one particle flying to infinity and eventually ending up in subsystem $R(t)$, while the other of the entangled pair falls into the black hole. Thus 
the radiation process generates more and more entanglement between the interior region of the black hole and $R(t)$,  
see Fig.~\ref{fig:BH}.
  
Let us now see how to calculate the entanglement entropies of $B(t)$ and $R(t)$ using the prescription~\eqref{hrn1} 
in the semi-classical regime of $\hbar G_N \to 0$. For $S_{B(t)}$ we take $A$ in~\eqref{hrn1} to be the whole boundary 
system and thus $\p A$ is empty. Before considering the quantum extremal surface for $S_{B(t)}$, let us first consider~\eqref{hrn}. The classical extremal surface $\ga$ is easy to find: consider all surfaces along the radial direction in Fig.~\ref{fig:BH}; clearly the one with the smallest area is the ``center'' of AdS, with zero area, and the corresponding entanglement wedge is the the whole AdS spacetime. Then we find that $S_{B(t)}$ is solely given by the second term in~\eqref{hrn}, and hence
\be \label{p1}
S_{B(t)} = S_{\rm AdS} = S_{R(t)} = \sS_{R (t)} \,,
\ee
where $\tilde  S_{\rm AdS}$ is the entanglement entropy of the full AdS spacetime and the second equality follows from realizing that 
the entanglement between the AdS spacetime and the radiation subsystem $R(t)$ is created through the Hawking pair creation process 
highlighted in the previous paragraph. In the last equality $\sS_{R(t)}$ is the coarse-grained thermodynamic entropy of the radiation; the equality follows from the fact that via the Hawking pair creation process $R(t)$ becomes maximally entangled, and that therefore 
the entanglement and thermodynamic entropies coincide. Before the Page time, this calculation in fact coincides with that using the quantum extremal surface~\eqref{hrn1}. The reason is very simple: at early times $S_{\rm AdS} = S_{R(t)} = \sS_{R (t)} $ is of order $(\hbar G_N)^0$, much smaller than any classical area contribution which is proportional to $(\hbar G_N)^{-1}$. Thus whether including the second term in~\eqref{hrn1} in the extremization process makes no difference, and 
the quantum extremal surface is still the one with the smallest classical area. 

When an $O(1)$ fraction of the black hole has evaporated, the entanglement between the interior of the black hole 
and the radiation becomes of order $(\hbar G_N)^{-1}$, the competition between the two terms in~\eqref{hrn1} becomes important. 
The location of the quantum extremal surface then requires an explicit calculation. 
One finds that after the Page time (i.e. when $\sS_{R(t)} > \sS_{\rm BH}$), the quantum extremal surface lies at a distance of order $O(G_N)$ inside the black hole horizon, as indicated by surface $K$ in Fig.~\ref{fig:BH}. The answer can be understood intuitively: the entanglement entropy for the region inside the horizon now becomes so large that pushing the surface to the left, the decrease in the classical area is more than compensated by the increase in the entanglement entropy. Pushing the surface to the right increases the classical area. 
One thus finds that 
\be \label{p2}
S_{B(t)} = {A(K) \ov 4 \hbar G_N} + S_{KB} \approx \sS_{\rm BH} + O((\hbar G_N)^0)
\ee
where $A(K)$ is the area of $K = \ga_{B(t)}$, the quantum extremal surface for $B(t)$, and  
$S_{KB}$ is the bulk entanglement entropy for the region $KB$, which is the entanglement wedge of $B(t)$. 
Note that $A(K) = A (H) + O(G_N)$ is approximately equal to the area of the horizon and thus the first term in~\eqref{p2} is 
approximately equal to the classical entropy of the black hole. 
The region $KB$ is weakly entangled with the radiation; its contribution to $S_{B(t)}$ comes from the usual entanglement entropy of graviton or matter fields which is of order $(\hbar G_N)^0$, and thus much smaller than the first term.

Combining~\eqref{p1} and~\eqref{p2} we see that $S_{B(t)}$ indeed follows the Page curve. 
To satisfy unitarity one still needs to show that $S_{R (t)} = S_{B(t)}$ after the Page time. Here one finds an interesting and potentially profound twist. Naively $S_{R(t)}$ is given by the number of entangled pairs created by the Hawking process and one finds $S_{R(t)}$ always increases, which would lead to a violation of unitarity and thus information loss. 
Unitarity implies there must be quantum correlations among radiation emitted at different times 
in order to have $S_{R (t)} = S_{B(t)}$ and to follow the Page curve. It turns out such correlations can be accounted for 
semi-classically if one treats the subregion duality seriously.  

From subregion duality, after the Page time, the boundary system $B(t)$ can only describe the region outside 
the quantum extremal surface, i.e. the region between $K$ and $B$ in Fig.~\ref{fig:BH}. 
Since radiation $R(t)$ is the complement of $B(t)$, again from subregion duality, the region inside the quantum extremal surface, denoted by $\sI$ in Fig.~\ref{fig:BH} must correspond to the entanglement wedge for $R(t)$. Then from~\eqref{hrn1}, 
in calculating $S_{R(t)}$ semi-classically we must include the contribution from $\sI (t)$, i.e. 
\be \label{k3}
S_{R(t)} = {A(K) \ov 4 \hbar G_N} +  S_{\sI \cup R (t)} = S_{B(t)} , 
\ee
with $K$ now regarded as the quantum extremal surface for the entanglement wedge of 
$R(t)$, i.e. $K = \p \sI = \ga_{R(t)}$,  and $S_{\sI \cup R (t)} =  S_{KB}$ from purity of the bulk state. 
Equation~\eqref{k3} can also be derived independently by assuming $\sI$ arbitrary and extremizing with respect to 
the variation of $\sI$~\cite{Almheiri:2019hni}. We note that $S(\sI \cup R (t)) $ is small since $R(t)$ is entangled with $\sI$.  
Such entanglement does not contribute to the entanglement entropy of their union.

The contribution from $\sI$ is highly unintuitive from the usual semi-classical perspective,  and is dubbed as an ``island'' contribution in~\cite{Almheiri:2019hni}, but as we saw, it is a direct consequence of subregion duality~\cite{Penington:2019npb}. It illustrates the power of subregion duality.

\section{Outlook} \label{sec:outlook}

In this review we sampled some recent exciting developments lying at the interface of quantum many-body physics, gravity, and quantum information. On the one hand, powerful methods from holography lead to new insights into quantum many-body dynamics in questions such as phases of strongly correlated systems and quantum many-body chaos. On the other hand, questions motivated from quantum many-body systems and quantum information lead to a deeper understanding of black holes as well as the fundamental structure of the duality. We expect synergies among these disciplines will continue to flourish, with many  more future developments. 

One particular area we expect could see significant development is using insights from holographic systems to guide experiments in strongly correlated electronic systems, 
including the search for hydrodynamic behavior 
new experimental probes of high temperature superconducting cuprates, and 
indications of highly entangled nature of strongly correlated phases.
See also~\cite{Zaanen:2018edk} for recent work advocating such efforts.
We mention in passing that insights from holographic duality have already been actively used to motivate experimental probes of QGPs created in the RHIC and LHC, see~\cite{Busza:2018rrf} for a recent review.   

Another area which could see further theoretical development is the use of
holography to study far-from-equilibrium dynamics, such as non-equilibrium steady states and thermalization. Non-equilibrium problems are notoriously difficult to deal with, not to mention at strong coupling and including quantum effects. Using holographic duality one can describe and follow the real time evolution of far-from-equilibrium systems, including those which are spatially inhomogeneous and anisotropic, by solving partial differential gravity equations (PDEs). Some important insights have already been obtained. See e.g.~\cite{CasalderreySolana:2011us} and~\cite{Liu:2018crr} for more recent progress.  

Finally further integrations of ideas from quantum many-body dynamics and quantum information 
should lead to more insights into the structure of spacetime, and help unlock mysteries of quantum gravity. 
After all, through the duality, all puzzles of quantum gravity can in principle be formulated in terms of quantum many-body dynamics without gravity. For example, one outstanding question is the fate of black hole or big bang type of singularities in 
quantum gravity theory.

\vspace{0.2in}   \centerline{\bf{Acknowledgements}} \vspace{0.2in}
We would like to thank Netta Engelhardt, Daniel Harlow,  Alexander Krikun, Joseph Minahan, and Jan Zaanen 
 for discussions. 
This work is supported by the Office of High Energy Physics of U.S. Department of Energy under grant Contract Number  DE-SC0012567. This work has also been supported by the Fonds National Suisse de la Recherche Scientifique (Schweizerischer Nationalfonds zur F\"orderung der wissenschaftlichen Forschung) through Project Grants 200021\_ 162796 and 200020\_ 182513 as well as the NCCR 51NF40-141869 The Mathematics of Physics (SwissMAP).

\bibliographystyle{utphys}
\bibliography{adscmtbib}{}
\end{document}